\documentclass[12pt]{article}       
\usepackage{latexsym}
\usepackage{eufrak}
\usepackage{euscript}
\usepackage{graphicx}

\title{Computer Model of  the Two-Pinhole Interference Experiment Using Two-Dimensional Gaussian Wave-Packets}

\author{ P. N. Kaloyerou\footnote{ and Wolfson College, Linton Road, Oxford OX2 6UD, UK.}  $\;$\& A. M. Ilunga \\
The University of Zambia\\ School of Natural Sciences \\
Department of Physics \\ 
Lusaka 10101\\ 
 Zambia}

\newcommand{\rb}{\mbox{$\vec{r}$}}
\newcommand{\pv}{\mbox{$\vec{p}$}}
\begin{document}



\maketitle

\begin{abstract}

The two-slit interference experiment has been modelled a number of times using Gaussian wave-packets and the Bohm-de Broglie causal interpretation. Here we consider the experiment with pinholes instead of slits and model the experiment in terms of two-dimensional Gaussian wave-packets and the Bohm-de Broglie causal interpretation.\\ \mbox{}\mbox{}\\
PACS NO. 03.65.-w 
\end{abstract}


\section{Introduction\label{intro}}
The  first computer models of quantum systems based on the Bohm-de Broglie causal interpretation \cite{B52,DEBR60} were developed by Dewdney in his  Ph.D thesis (1983) \cite{DEWD83}. These models include the two-slit experiment and scattering from square barriers and square wells.  Some of the results appeared in earlier articles  with Phillipides, Hiley (1979)  \cite{DEWD79} and with  Hiley (1982) \cite{DEWD82}. In later years,  Dewdney developed a computer model of Rauch's Neutron interferometer (1982) \cite{DEWD85} and,  with  Kyprianidis and Holland,  models of  a spin measurement in a Stern-Gerlach experiment (1986) \cite{DEWD86}. He also went on, with  Kyprianidis and Holland, to develop computer models of spin superposition in neutron interferometry   (1987) \cite{DEWD87} and  of Bohm's spin version of the Einstein, Rosen, Podolsky experiment (EPR-experiment) (1987) \cite{DEWDEPR87}. A review of this work appears in a 1988 {\it Nature} article  \cite{DEWD88}. The computer models of spin were based on the 1955 Bohm-Schiller-Toimno causal interpretation of spin \cite{BST55}. Home and Kaloyerou in 1989 reproduced the computer model of the two-slit interference experiment \cite{K89} in the context of arguing against  Bohr's Principle of Complementarity \cite{BR28}. 

Though computer models of the two-slit experiment with each slit modeled by a  one-dimensional Gaussian wave-packet have existed for many years, the extension to pinholes has never been made. We thought, therefore, that it might be interesting to attempt such an extension by modeling each pinhole by a two-dimensional Gaussian wave-packet. Though no new conceptual results are expected, we thought it might be interesting to see if the trajectories, now in three dimensional space, retain the characteristic features of the two-slit case. We shall see that a quantum potential structure is produced which guides the particles to the bright fringes as in the two-slit case. With the pinholes along the $x$-axis, trajectories in the $xy$-plane  (see Fig. \ref{OrAxes}) show the same interference behaviour as in the two-slit case, while trajectories in the $zy$-plane show no interference.

\section{The mathematical model\label{MM}}
Phillipides et al \cite{DEWD79} derived the Gaussian function they used to model each of the two slits in the two-slit experiment using Feynman's path integral formulation. We, instead, have generalised the one dimensional Gaussian solutions of the Schr\"{o}dinger equation developed by Bohm in chapter three of his book \cite{B51} to two-dimensions. Phillipides et al \cite{DEWD79} considered Young's two-slit experiment for electrons and used the values of an actual  Young's two-slit experiment for electrons performed by J\"{o}nsson in 1961 \cite{jon61}. We will also model the interference of electrons and use  J\"{o}nsson's values, except that we will vary slightly the distance between the pinholes and the detecting screen in order to obtain clearer interference or quantum potential plots. We will, in any case,  give the values used for each case we consider. 

\begin{figure}[h]
\unitlength=1in
\hspace*{1.3in}\includegraphics[width=3.4in,height=1.8in]{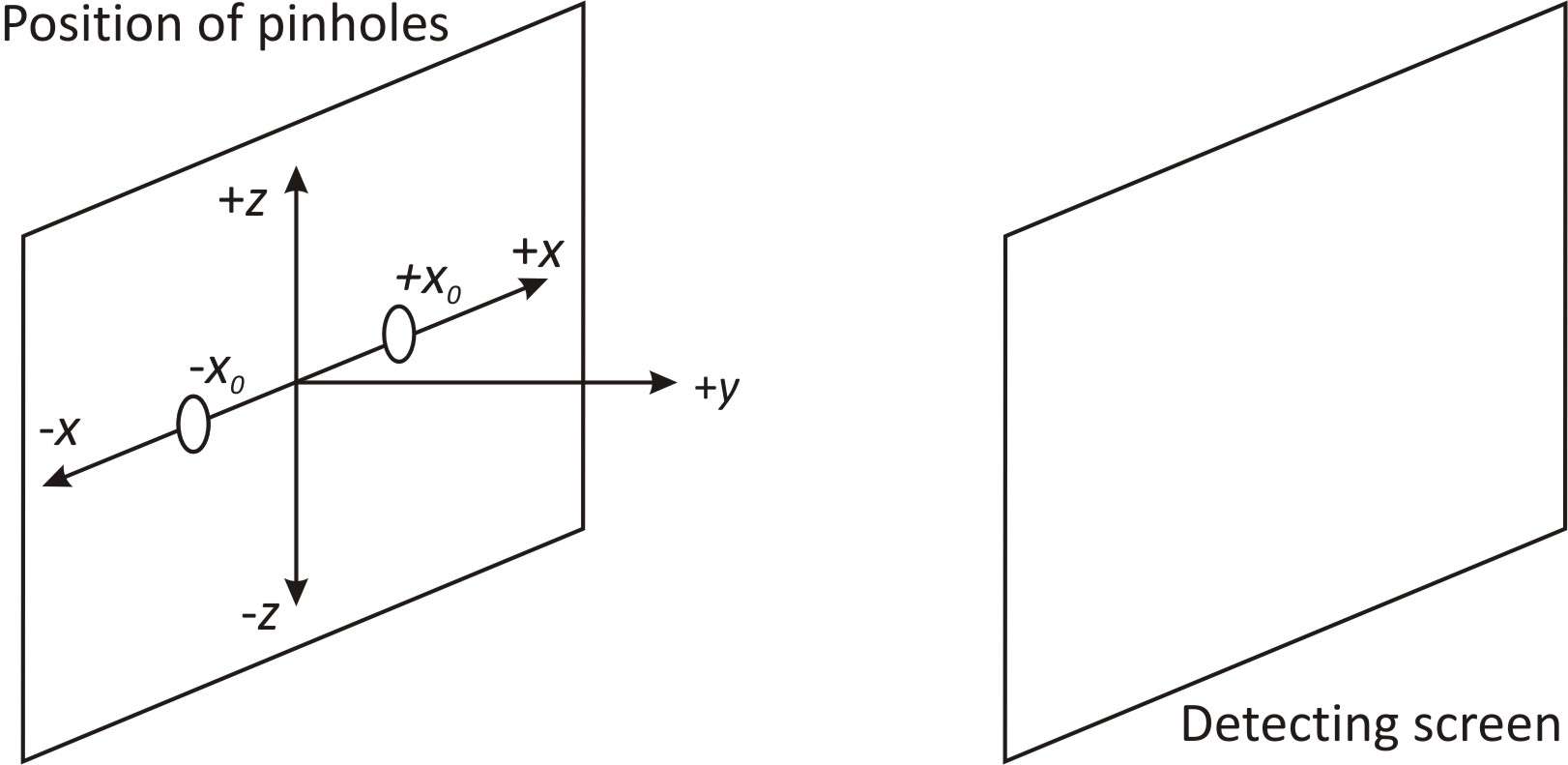}  
\caption{The orientation of the axes.\label{OrAxes}}
\end{figure}

\begin{figure}[h]
\unitlength=1in
\hspace*{1.3in}\includegraphics[width=3.4in,height=1.8in]  {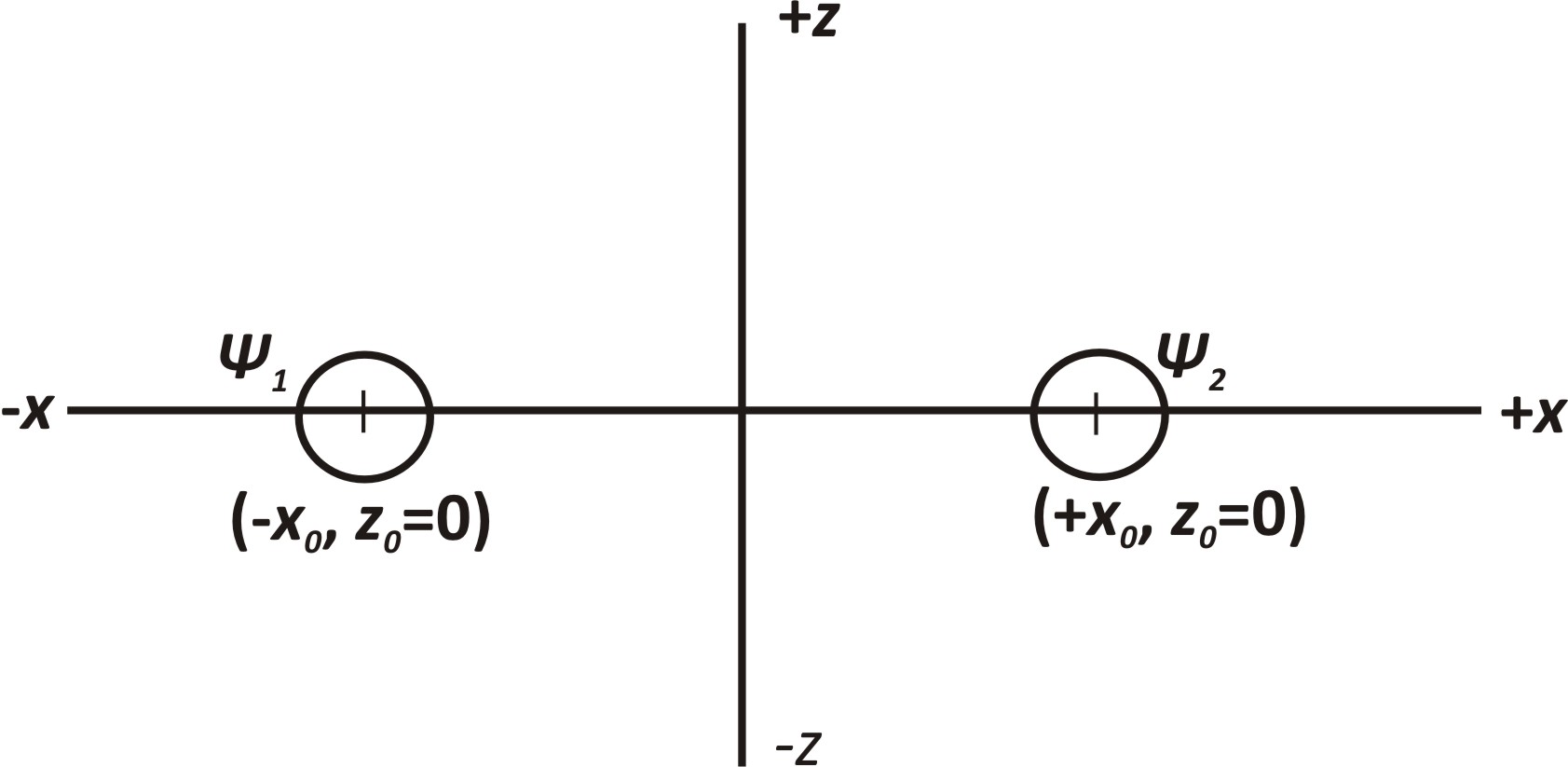}  
\caption{Positions of the pinholes: The two pinholes, represented by the two-dimensional Gaussian wave-packets $\psi_1$ and $\psi_2$, are placed at $x_0=\pm5\times10^{-7}$ m and  $z_0=0$ m. The width of  the Gaussian packet on the negative $x$-side is $\Delta x_{n0}=\Delta z_{n0}=7\times 10^{-8}$ m, while the width  of the Gaussian packet on the positive $x$-side is $\Delta x_{p0}=\Delta z_{p0}=\Delta x_{n0}$  or $\Delta x_{p0}=\Delta z_{p0}=2\Delta x_{n0}$ for the case of unequal widths.  The velocity components are  $v_x=150\;\mathrm{ms}^{-1}$, $v_y=1.3\times 10^{8}\;\mathrm{ms}^{-1}$ and  $v_z=0\;\mathrm{ms}^{-1}$.\label{figPos} }
\end{figure}

The orientation of the axes and the position of the pinholes are shown in Figs. \ref{OrAxes} and \ref{figPos}, respectively.
The pinholes are represented by two dimensional Gaussian wave-packets  $\psi_1$ and $\psi_2$ given by
\begin{eqnarray}
\psi_1(x,y,z,t)&=&A_1\tilde{R}_{1}\nonumber\\
&&\times\exp\left[{-  \frac{(x+x_0-v_x t)^2}{2\Delta x_n^2}}\right]\exp\left[{\frac{i\alpha t(x+x_0-v_x t)^2}{2\Delta x_{n1}^2}}\right]\nonumber\\
&&\times\exp\left[{-  \frac{(z+z_0-v_z t)^2}{2\Delta z_n^2}}\right] \exp\left[{\frac{i\alpha t(z+z_0-v_z t)^2}{2\Delta z_{n1}^2}}\right]\exp\left[ik_x(x+x_0)\right]\nonumber\\
&&\times\exp\left[ik_z(z+z_0)\right]\exp\left(ik_y y\right)\exp\left[-i(\omega_x+\omega_z) t\right],\label{psi1}
\end{eqnarray}
\newpage
\mbox{}\\
and
\begin{eqnarray}
\psi_2(x,y,z,t)&=&A_2\tilde{R}_{2}\nonumber\\
&&\times\exp\left[{-  \frac{(x-x_0+v_x t)^2}{2\Delta x_p^2}}\right]\exp\left[{\frac{i\alpha t(x-x_0+v_x t)^2}{2\Delta x_{p1}^2}}\right]\nonumber\\
&&\times\exp\left[{-  \frac{(z-z_0+v_z t)^2}{2\Delta z_p^2}}\right] \exp\left[{\frac{i\alpha t(z-z_0+v_z t)^2}{2\Delta z_{p1}^2}}\right]\exp\left[-ik_x(x-x_0)\right]\nonumber\\
&&\times\exp\left[-ik_z(z-z_0)\right]\exp\left(ik_yy\right)\exp\left[-i(\omega_x+\omega_z) t+i\chi)\right].\label{psi2}
\end{eqnarray}
The various functions and constants used are given by:
\begin{eqnarray}
&&\!\!\!\!\!\!\!\!\!\!\!\!\!\tilde{R}_{1}=\cos^2\frac{\pi}{4},\;\;\;\tilde{R}_{2}=\sin^2\frac{\pi}{4}\;\;\;\mathrm{(for\; equal\; amplitudes)},\nonumber\\
\alpha&=&\frac{\hbar}{m},\; \;\;v_x=\frac{\hbar k_x}{m}, \;\;\;\omega_x=\frac{\hbar k_x^2}{2m},\;\;\;v_z=\frac{\hbar k_z}{m}, \;\;\;\omega_z=\frac{\hbar k_z^2}{2m},\nonumber\\     \nonumber\\
&&\!\!\!\!\!\!\!\!\!\!\!\!\!\Delta x_{n0}=\mathrm{width \;of\; the}\;-x_0\;\mathrm{wavepacket}, \;\;\;\Delta z_{n0}=\Delta x_{n0}\nonumber\\
&&\!\!\!\!\!\!\!\!\!\!\!\!\!\Delta x_{p0}=\mathrm{width \;of\; the}\;+x_0\;\mathrm{wavepacket}, \;\;\;\Delta z_{p0}=\Delta x_{p0}\nonumber\\
&&\!\!\!\!\!\!\!\!\!\!\!\!\!A_1(t)=A_{xn}(t)A_{zn}(t)= \left(    \frac{2\pi}{\Delta x_{n0}^2+i\alpha t}   \right)^{\frac{1}{2}}\left(\frac{2\pi}{\Delta z_{n0}^2+i\alpha t}\right)^{\frac{1}{2}} \nonumber \\ \nonumber\\
&&\;\;=\beta_{xn} (t)e^{i\theta_{xn}(t)}\beta_{zn}(t)e^{i\theta_{zn}(t)}=\beta_1e^{ i2\theta_1}, \label{AA11}\\
&&\!\!\!\!\!\!\!\!\!\!\!\!\!A_2(t)=A_{xp}(t)A_{zp}(t)= \left(    \frac{2\pi}{\Delta x_{p0}^2+i\alpha t}   \right)^{\frac{1}{2}}\left(\frac{2\pi}{\Delta z_{p0}^2+i\alpha t}\right)^{\frac{1}{2}}  \nonumber\\ \nonumber\\
&&\;\;=\beta_{xp} (t)e^{i\theta_{xp}(t)}\beta_{zp}(t)e^{i\theta_{zp}(t)}=\beta_2 e^{ i2\theta_2}, \label{AA22}
\end{eqnarray}
where
\begin{eqnarray}
&&\!\!\!\!\!\!\!\!\!\!\!\!\!\beta_{xn}(t)=\beta_{zn}(t) =\left( \frac{4\pi^2}{\Delta x_{n0}^4+\alpha^2 t^2}\right)^{\frac{1}{4}},\;\;\;\beta_{1}(t)=\left( \frac{4\pi^2}{\Delta x_{n0}^4+\alpha^2 t^2}\right)^{\frac{1}{2}}\nonumber\\ \nonumber\\
&&\!\!\!\!\!\!\!\!\!\!\!\!\!\theta_1=\theta_{xn}(t)=\theta_{zn}(t)  =\frac{1}{2} \tan^{-1} \left(-\frac{\alpha t}{\Delta x_{n0}^2}  \right)+2k\pi,\;\;\;k=0,1,\ldots,\nonumber\\ \nonumber\\
&&\!\!\!\!\!\!\!\!\!\!\!\!\!\beta_{xp}(t)=\beta_{zp}(t) =\left( \frac{4\pi^2}{\Delta x_{p0}^4+\alpha^2 t^2}\right)^{\frac{1}{4}},\;\;\;\beta_{2}(t)=\left( \frac{4\pi^2}{\Delta x_{p0}^4+\alpha^2 t^2}\right)^{\frac{1}{2}}\nonumber\\ \nonumber\\
&&\!\!\!\!\!\!\!\!\!\!\!\!\!\theta_2=\theta_{xp}(t)=\theta_{zp}(t)  =\frac{1}{2} \tan^{-1} \left(-\frac{\alpha t}{\Delta x_{p0}^2}  \right)+2k\pi,\;\;\;k=0,1,\ldots,\nonumber\\ \nonumber\\
&&\!\!\!\!\!\!\!\!\!\!\!\!\!\!\Delta x_n^2=\Delta z_n^2 =\left(\Delta x_{n0}^2+\frac{\alpha^2 t^2}{\Delta x_{n0}^2}\right),\;\;\;\Delta x_{n1}^2=\Delta z_{n1}^2=\left(\Delta x_{n0}^4+\alpha^2 t^2\right),\nonumber\\
&&\!\!\!\!\!\!\!\!\!\!\!\!\!\!\Delta x_p^2= \Delta z_p^2 =\left(\Delta x_{p0}^2+\frac{\alpha^2 t^2}{\Delta x_{p0}^2}\right),\;\;\;\Delta x_{p1}^2=\Delta z_{p1}^2=\left(\Delta x_{p0}^4+\alpha^2 t^2\right),\nonumber
\end{eqnarray}
Further definitions and values of quantities used in the plots are given in Table \ref{COMPP}.1.   Note that  the Gaussian wave packets are functions of $x$ and $z$, while the $y$-behaviour is represented by a plane wave. Plane  waves are useful  idealisations that are not realisable in practice. This leads to a computer model in which the intensity $R^2$ and quantum potential $Q$ at a given time maintain the same form from $y=-\infty$ to $+\infty$.	However, both the intensity and quantum potential evolve in time so that an electron at each position of its trajectory sees an evolving intensity and quantum potential. 

The total wave function is the sum of the two wave packets:
\begin{equation}
\psi=\psi_1+\psi_2
\end{equation}

Our computer model is based on the Bohm-de Broglie causal interpretation and we refer the reader to Bohm's original papers for details of the interpretation \cite{B52}. Here we will give only a very brief outline in order to introduce the elements we will need to develop the formulae and equations needed for the model. The interpretation is obtained by substituting $\phi(x,y,z,t)=R(x,y,z,t)\exp(iS(x,y,z,t/\hbar)$, where $R$ and $S$ are two fields which codetermine one another, into the Schr\"{o}dinger equation
\[
i\hbar\frac{\partial \psi}{\partial t}=-\frac{\hbar^2}{2m}\nabla^2\psi+V\psi,
\]
where $V=V(x,y,z,t)$. Differentiating and equating real and imaginary terms gives two equations. One is the usual  continuity equation,
\begin{equation}
\frac{\partial R^{2}}{\partial t}+\nabla\cdot\left(R^{2}\frac{\nabla S}{m}\right)=0,
\end{equation}
which expresses the conservation of probability $R^2$. The other is a Hamilton-Jacobi type equation
\begin{equation}
 -\frac{\partial S}{\partial t}= \frac{(\nabla S)^{2}}{2m}+V+\left    (-\frac{\hbar^{2}}{2m}\frac{\nabla^{2} R}{R}\right).
\end{equation}
This differs from the classical  Hamilton-Jacobi equation by the extra term
\begin{equation}
Q=-\frac{\hbar^{2}}{2m}\frac{\nabla^{2} R}{R},
\end{equation}
which Bohm called the quantum potential. The classical  Hamilton-Jacobi equation describes the behaviour of a particle with energy $E$, momentum  $p$ and velocity  $v$ under the action of a potential $V$, with the energy, momentum and velocity given by
\begin{eqnarray}
E&=&-\frac{\partial S}{\partial t}, \nonumber\\
p&=&\nabla S,\nonumber\\
v_p(\rb)&=&\frac{d \rb}{dt}=\frac{\nabla S}{m}.  \label{ENMOM}
\end{eqnarray}
Bohm retain's these definitions, while de-Broglie focused the third definition and called it the guidance  formula. This allows quantum entities such as electrons, protons, neutrons etc. (but not photons\footnote{The causal interpretation based on the Schr\"{o}dinger is obviously non-relativistic, but it is more than adequate for the description of the behaviour of electrons, protons, neutrons etc., in a large range of circumstances. This is not so for photons, the proper description of which requires quantum optics, which is based on the second-quantisation of Maxwell's equation.  Photons, more generally, the electromagnetic field are described by the causal interpretation of boson fields, which includes the electromagnetic field \cite{K85}.}) to be viewed as particles (always) with energy, momentum and velocity given by (\ref{ENMOM}). Particle trajectories are found by integrating $v(\rb)$ given in (\ref{ENMOM}). The extra $Q$ term produces quantum behaviour such as the interference of particles (which is what we will model in this article). Strictly, since the $R$ and $S$-fields codetermine one another, the $S$-field is as much responsible for quantum behaviour as the $R$-field; the $S$-field through the guidance formula, and the $R$-field through the quantum potential.  

The Born probability rule is an essential interpretational element that links theory with experiment. As such it will remain a part  of any interpretation of the quantum theory. This is certainly true for the causal interpretation, where probability enters because the initial positions of particles cannot be determined precisely. Instead, initial positions are given with a probability found from the usual probability density $|\psi(x,y,z,t=0)|^2=R(x,y,z,t=0)^2$.

The results of the usual interpretation are identical with those of the causal interpretation as long as the following assumptions are satisfied:
         \begin{enumerate}
               \item[(1)] The  $\psi$-field satisfies Schr\"{o}dinger equation.
          \item[(2)]  Particle momentum is restricted to $\pv=\nabla S$.
               \item[(3)]   Particle position at time $t$ is given by the
                probability density $|\psi(\rb,t)|^2$.
         \end{enumerate}

To obtain the intensity, $Q$ and trajectories we must first find the $R$ and $S$-fields defined by $\psi=Re^{iS/\hbar}$ in terms of $R_1$, $R_2$, $S_1$ and $S_2$ defined by $\psi_1=A_1 R_1e^{iS'_1/\hbar}  =\beta_1R_1e^{iS_1/\hbar}$ and $\psi_2=A_2 R_2e^{iS'_2/\hbar} =\beta_2 R_2e^{iS_2/\hbar}$. We do this by first noting Eq. (\ref{AA11}) for $A_1$ and Eq. (\ref{AA22}) for $A_2$ and then comparing  $\psi_1=A_1 R_1e^{iS'_1/\hbar}  =\beta_1R_1e^{iS_1/\hbar}$ and $\psi_2=A_2 R_2e^{iS'_2/\hbar} =\beta_2 R_2e^{iS_2/\hbar}$ with Eqs. (\ref{psi1}) and (\ref{psi2}) to  get
\begin{eqnarray}
R_1(x,z,t)&=& \beta_1 \tilde{R}_{1}\exp\left[{-\frac{(x+x_0-v_x t)^2}{2\Delta x_n^2}}\right]\exp\left[{-  \frac{(z+z_0-v_z t)^2}{2\Delta z_n^2}}\right] \nonumber\\
R_2(x,z,t)&=& \beta_2 \tilde{R}_{2}\exp\left[{-\frac{(x-x_0+v_x t)^2}{2\Delta x_p^2}}\right] \exp\left[{- \frac{(z-z_0+v_z t)^2}{2\Delta z_p^2}} \right]      \nonumber\\
S_1(x,y,z,t)&=&{\frac{\hbar\alpha t(x+x_0-v_x t)^2}{2\Delta x_{n1}^2}}+{\frac{\hbar\alpha t(z+z_0-v_z t)^2}{2\Delta z_{n1}^2}}\nonumber\\
&& + \hbar k_x(x+x_0)+\hbar k_z(z+z_0)+\hbar k_y y-\hbar(\omega_x+\omega_z) t+2\hbar\theta_1\nonumber  \nonumber\\
S_2(x,y,z,t)&=&\frac{\hbar\alpha t(x-x_0+v_x t)^2}{2\Delta x_{p1}^2}+\frac{\hbar\alpha t(z-z_0+v_z t)^2}{2\Delta z_{p1}^2}\nonumber\\
&&-\hbar k_x(x-x_0)-\hbar k_z(z-z_0)+\hbar k_y y-\hbar(\omega_x+\omega_z) t+\hbar\chi+2\hbar\theta_2.  \nonumber\\
\end{eqnarray}
The intensity (probability density) is easily found from $|\psi|^2=R^2$:
\begin{equation}
R^2= R_1^2+R_2^2+2R_1^2R_2^2\cos\left( \frac{S_1-S_2}{\hbar}\right).\label{RSQ}
\end{equation}
The quantum potential ($Q$) is found from:
\begin{eqnarray}
Q&=&-\frac{\hbar^{2}}{2m}\frac{\nabla^{2} R}{R}\nonumber\\
&=&-\frac{\hbar^{2}}{2m}\frac{1}{R}\left(\frac{\partial^2  R}{\partial x^2}+   \frac{\partial^2  R}{\partial y^2}+ \frac{\partial^2  R}{\partial z^2}  \right)=Q_x+Q_y+Q_z,\nonumber
\end{eqnarray}
where
\begin{equation}
Q_x=-\frac{\hbar^{2}}{2mR}\frac{\partial^2  R}{\partial x^2}=\frac{\hbar^{2}}{8mR^4}\left(\frac{\partial  R^2}{\partial x}\right)^2 -\frac{\hbar^{2}}{4mR^2}\frac{\partial^2  R^2}{\partial x^2}, \label{QPF}
\end{equation}
with similar formulae for $Q_y$ and $Q_z$. Substituting Eq. (\ref{RSQ}) into the formulae for $Q_x$ and $Q_y$  and differentiating gives $Q_y=0$ and
\begin{eqnarray}
Q_x&=&- \frac{\hbar^2}{4mR^4} \left[ \frac{(x+x_0-v_x t)}{\Delta x_n^2}R_1^2+\frac{(x-x_0+v_x t)}{\Delta x_p^2}R_2^2          \right.        \nonumber\\
&&+ \left.  R_1^2 R_2^2 \left[     \left(  \frac{(x+x_0-v_x t)}{\Delta x_n^2}+\frac{(x-x_0+v_x t)}{\Delta x_p^2}  \right)  \cos(S_{12}) +S_{x12}\sin(S_{12})
      \right]\right]^2\nonumber\\
&&\!\!\!\!\!\!\!\!\!\!-\frac{\hbar^2}{2mR^2} \left[           R_1^2      \left(    \frac{2(x+x_0-v_x t)^2}{\Delta x_n^4}-\frac{1}{\Delta x_n^2}    \right) +  R_2^2\left(\frac{2(x-x_0+v_x t)^2}{\Delta x_p^4}-\frac{1}{\Delta x_p^2}    \right)           \right]    \nonumber\\
&&\;\;-\frac{\hbar^2R_1 R_2}{2mR^2}    \left(  \frac{(x+x_0-v_x t)^2}{\Delta x_n^4}-\frac{1}{\Delta x_n^2}+2\frac{(x+x_0-v_x t)(x-x_0+v_x t)}{\Delta x_n^2 \Delta x_p^2}  \right. \nonumber\\
&&\;\;\;\;\;\;\;\;\;\;\;\;\;\;\;\;\;\;\;\;\;\;\;\;\;\;\;\;\; \;\;\;\;\;\;\;\;\;\;\;\;\;\;\;\;\;\;\;\;\;\;\;\;\;+ \left. \frac{(x-x_0+v_x t)^2}{\Delta x_p^4}     -      \frac{1}{\Delta x_p^2}       \right)\cos(S_{12}) \nonumber\\
&&\;\,-\frac{\hbar^2R_1 R_2}{2mR^2}    \left[  2\left(   \frac{(x+x_0-v_x t)}{\Delta x_n^2}+\frac{(x-x_0+v_x t)}{\Delta x_p^2}  \right)S_{x12} \sin(S_{12})    \right. \nonumber\\
&&\;\;\;\;\;\;\;\;\;\;\;\;\;\;\;\;\;\;\;\;\;\;\;\;\;\;\;\;\;\;\;\;\;\;\;\;\;\;\;\;\;\;\;\;\;\;\;\;\;\;\;\;  \left. - S_{xx12}\sin(S_{12}) - S_{x12}^2 \cos(S_{12})               \right]\label{QPX}
\end{eqnarray}
where 
\begin{eqnarray}
&&\!\!\!\!\!\!\!\!\!\!\!\!\! S_{12}=S_1-S_2 \nonumber\\
&&\!\!\!\!\!\!\!\!\!\!\!\!\! S_{x12}=  \frac{\alpha t(x+x_0-v_x t)}{\Delta x_{n1}^2}  - \frac{\alpha t(x-x_0+v_x t)}{\Delta x_{p1}^2} + 2k_x\nonumber\\
&&\!\!\!\!\!\!\!\!\!\!\!\!\! S_{xx12}=  \frac{\alpha t}{\Delta x_{n1}^2}  - \frac{\alpha t}{\Delta x_{p1}^2}\nonumber
\end{eqnarray}
The formulae for $Q_z$ is identical to that of $Q_x$, except that $x$ is replaced by $z$ everywhere it appears.

The trajectories, as we have said, are found by integrating Eq. (\ref{ENMOM}). Therefore, to find the trajectories we do not need to find $S$,  only its derivatives with respect to $x,y,z$. This  can be done using  the formula
\begin{equation}
\nabla S=\frac{\hbar}{2i}\left( \frac{\nabla \psi}{\psi}-\frac{\nabla \psi*}{\psi*}    \right).
\end{equation}
We  get
\begin{equation}
\frac{\partial S}{\partial y}=\hbar k_y,
\end {equation}
and
\begin{eqnarray}
\frac{\partial S}{\partial x}&=&\frac{\hbar}{R^2}\left[    \frac{x}{\Delta x_n^2\Delta x_p^2 }R_1 R_2\sin(S_{12}) (\Delta x_n^2-\Delta x_p^2)          \right.                                               \nonumber\\
&&+  \frac{\alpha t x}{\Delta x_{n1}^2\Delta x_{p1}^2}\left[ R_1^2\Delta x_{p1}^2 +   R_2^2\Delta x_{n1}^2 + R_1 R_2\cos(S_{12})   \right](\Delta x_{n1}^2+\Delta x_{p1}^2)                                                \nonumber\\
&&\;-   \frac{(x_0-v_xt)}{\Delta x_n^2\Delta x_p^2 } R_1 R_2\sin(S_{12}) (\Delta x_n^2+\Delta x_p^2)                                                      \nonumber\\
&&\;+ \frac{(\alpha t x_0-\alpha t^2v_x)}{\Delta x_{n1}^2\Delta x_{p1}^2}\left[R_1^2\Delta x_{p1}^2 -  R_2^2\Delta x_{n1}^2 + R_1 R_2\cos(S_{12})   \right](\Delta x_{p1}^2-\Delta x_{n1}^2)                                                         \nonumber\\
&&\;\;\;+ k_x(  R_1^2 -  R_2^2)                                                 \nonumber\\
\end{eqnarray}
The $z$-derivative $\frac{\partial S}{\partial z}$ is identical to $\frac{\partial S}{\partial x}$, except that $x$ is everywhere replaced by $z$. From Eq. (\ref{ENMOM}),
\begin{equation}
v_p=v_{px}\hat{i}+v_{py}\hat{j}+v_{pz}\hat{k}=\frac{dx}{dt}\hat{i}+\frac{dy}{dt}\hat{j}+\frac{dz}{dt}\hat{k}=\frac{1}{m}\left(    \frac{\partial S}{\partial x}\hat{i}+\frac{\partial S}{\partial y}\hat{j}+\frac{\partial S}{\partial z}\hat{k}  \right),
\end{equation}
we see that to obtain the electron trajectories  $\rb(t)=x(t)\hat{i}+y(t)\hat{j}+z(t)\hat{k}$, we must solve the following differential equations with various initial conditions:
\begin{eqnarray}
\frac{dx(t)}{dt} =\frac{1}{m}\frac{\partial S}{\partial x}, \label{DXDT}\\
\frac{dy(t)}{dt} =\frac{1}{m}\frac{\partial S}{\partial y}, \label{DYDT}\\
\frac{dz(t)}{dt} =\frac{1}{m}\frac{\partial S}{\partial z}. \label{DZDT}
\end{eqnarray}
Note that the components of the particle veloctiy $v_p$  are different from the velocities of the wave packets $v_x$, $v_y$ and $v_z$. Eq. (\ref{DYDT}) can be solved immediately to give $y(t)=\hbar k_y t$. Eqs. (\ref{DXDT}) and (\ref{DZDT}) are coupled non-linear differential equations. These were solved numerically using  a Fortran program we wrote based on an adapted fourth-order Runge-Kutta algorithm \cite{BF89} with fixed step size. This completes the various elements  of he mathematical model. In the following section we show the various plots.  
 \section{The Computer plots\label{COMPP}}
\begin{figure}[h]
\unitlength=1in
\hspace*{0.7in}\includegraphics[width=4.6in,height=2.3in]{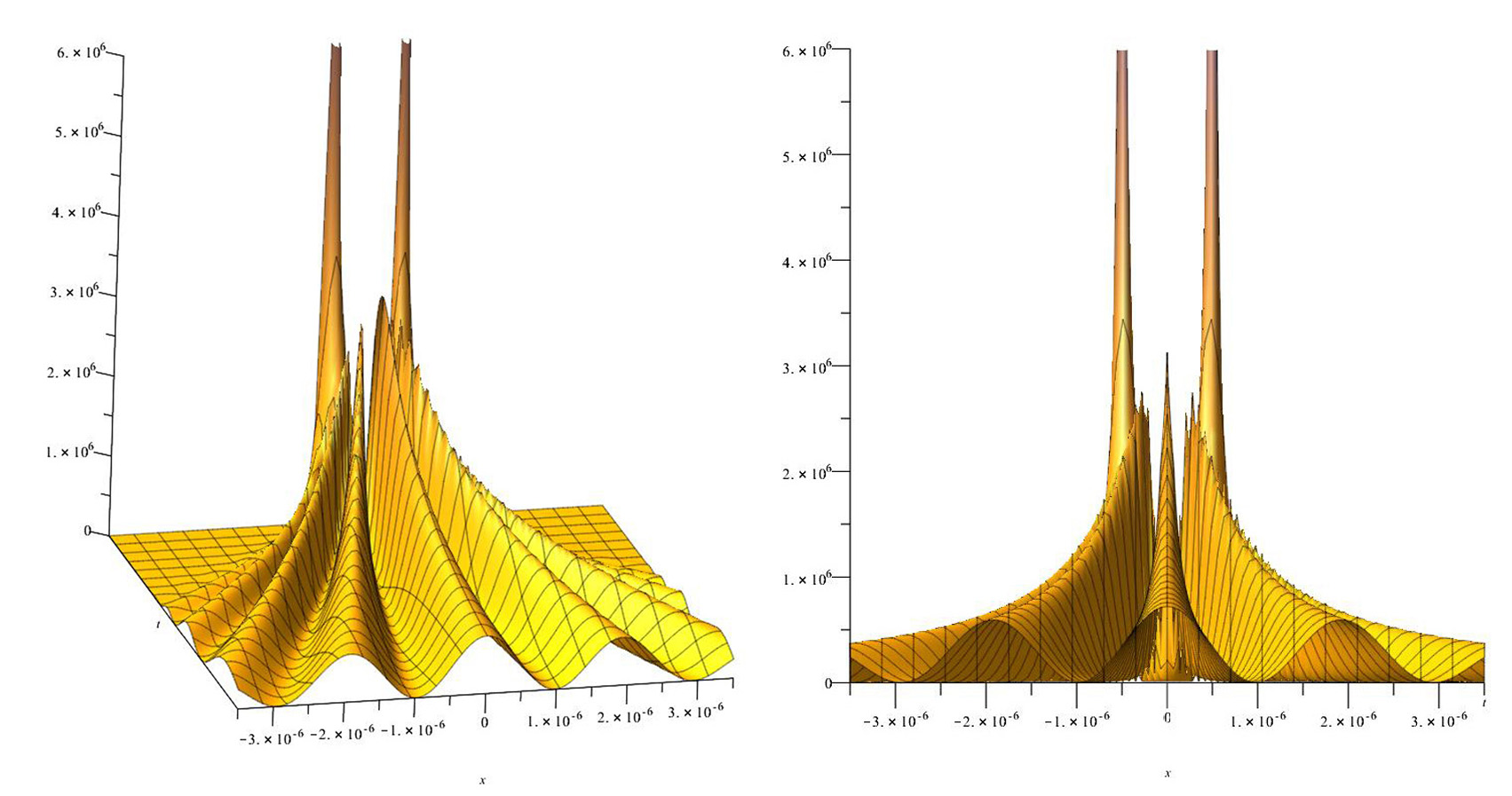}
\caption{Two orientations of the intensity in a two-slit interference  experiment modeled by  one-dimensional Gaussian wave-packets.\label{INT1DG}}
\end{figure}

\begin{figure}[h]
\unitlength=1in
\hspace*{0.7in}\includegraphics[width=4.6in,height=2.3in]  {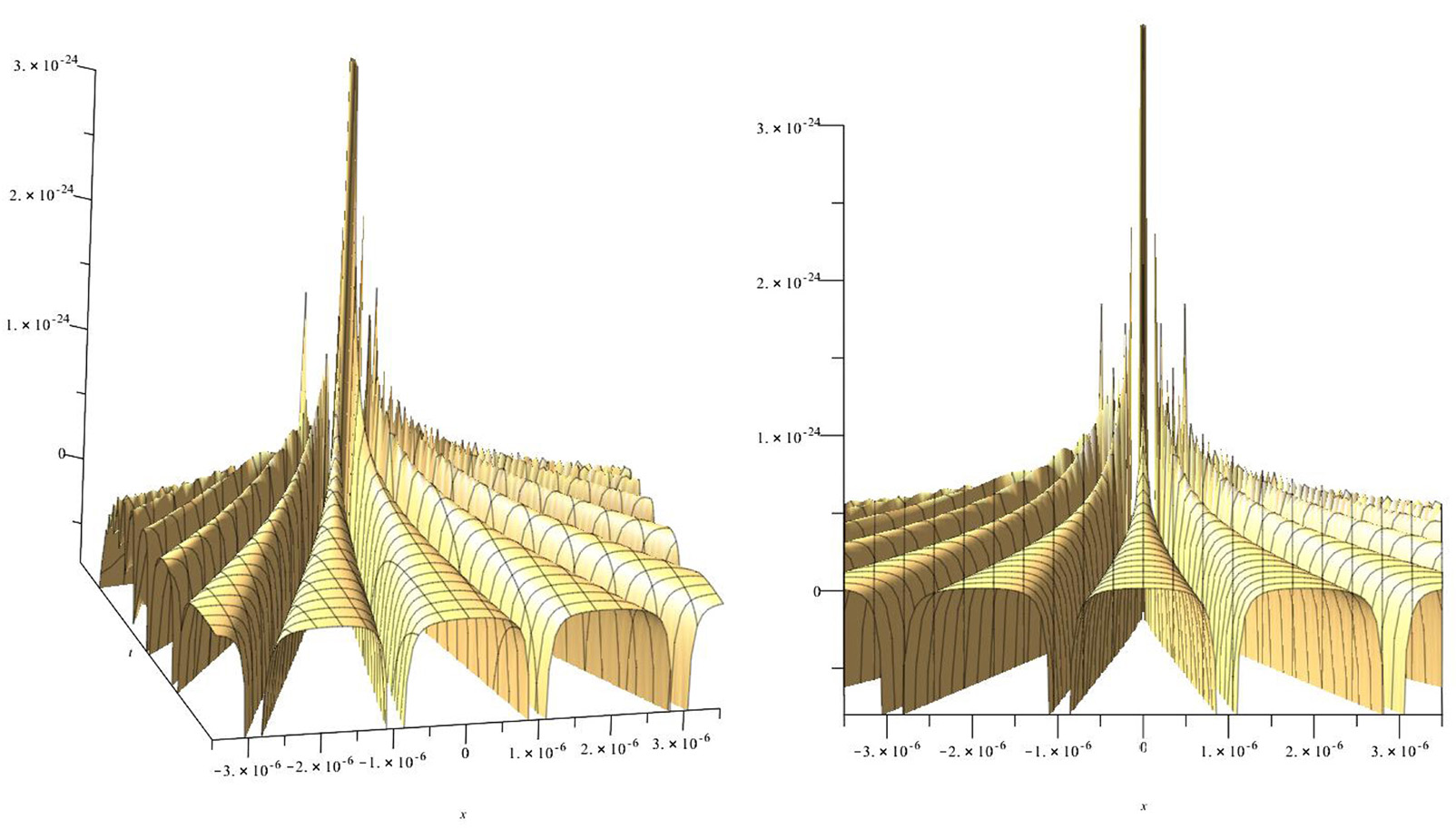}  
\caption{Two orientations of the quantum potential in a two-slit  interference experiment modeled by  one-dimensional Gaussian wave-packets.\label{QP1DG}}
\end{figure}

\begin{figure}[h]
\unitlength=1in
\hspace*{1.4in}\includegraphics[width=3in,height=2.6in]  {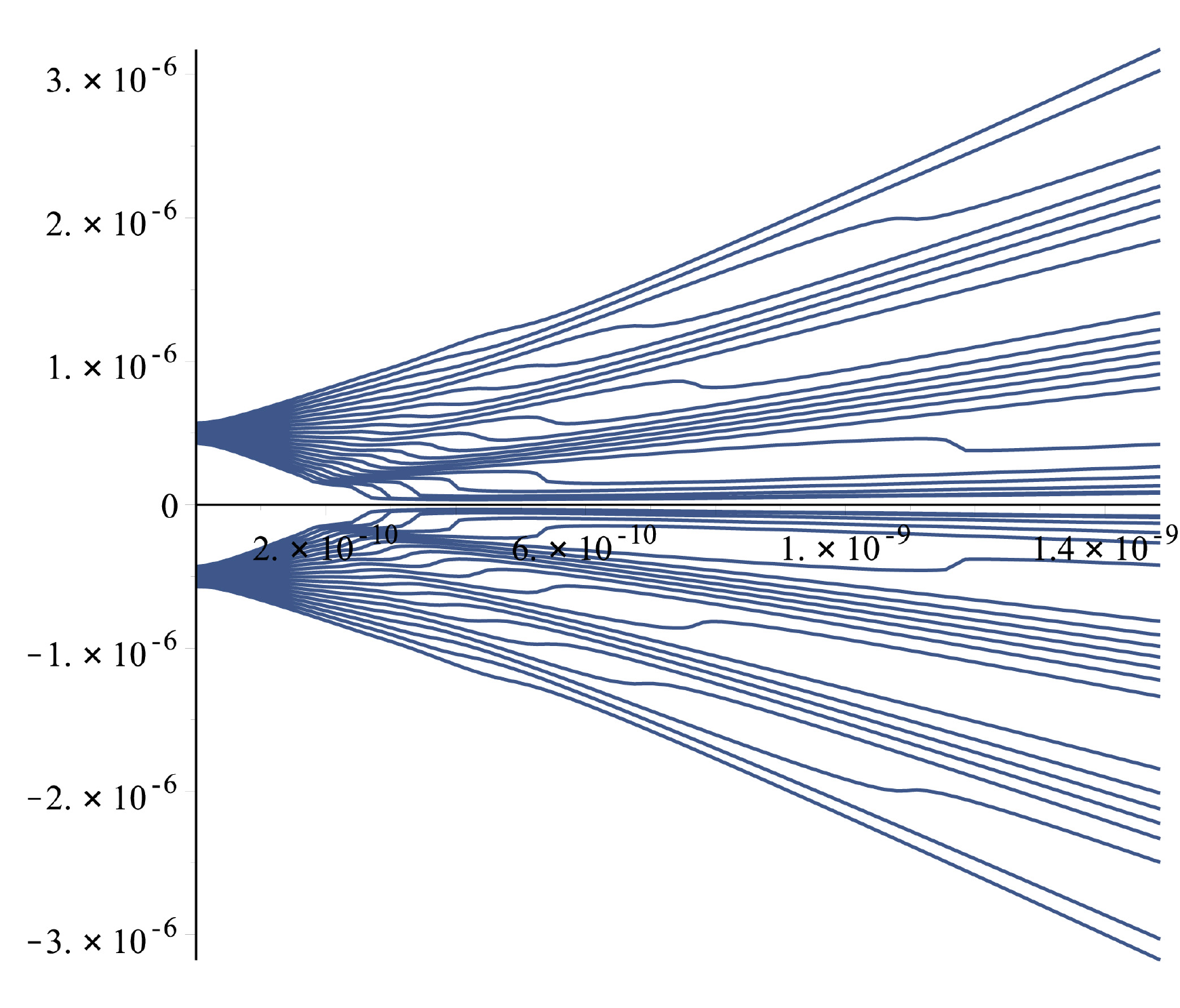}  
\vspace*{0.0in}
\caption{The Trajectories in a two-slit interference experiment modeled by a one-dimensional Gaussian wave-packet.\label{TRAJ1DG}}
\end{figure}

For the sake of comparison, we first reproduce plots of the intensity, $Q$ and trajectories for the two-slit experiment modeled by one-dimensional Gaussian wave-packets. These are shown in Figs. \ref{INT1DG},  \ref{QP1DG}, and  \ref{TRAJ1DG}. \\ \mbox{}\\
\begin{center}
\begin{tabular}{|l|} \hline 
\hspace*{.1in}{\bf Table \ref{COMPP}.1}   Definition and values of the quantities used in the plots. \hspace*{0.4cm}\\ 
\end{tabular}
\begin{tabular}{|l|l|l|} \hline\hline 
\textbf{Quantity}&\hspace*{.8in}\textbf{Definition}                &  \textbf{Value}   \\   \hline\hline  
$b$                     & Angle for equal amplitudes  &  $\pi/4$\\ \hline
$b$                     & Angle for unequal amplitudes  &  $\arccos(1/\sqrt{4})$ \\ \hline
$\tilde{R_1}$&   Amplitude   of $\psi_1$                    &$\cos^2 (b)$ \\  \hline
$\tilde{R_2}$&   Amplitude   of $\psi_2$                       &$\sin^2 (b)$\\ \hline
$x_0$ & $x$-distance of the center of the&   $5\times 10^{-7}$ m\\ 
                     &   pinhole from the origin        &  \\ \hline
$z_0$ & $z$-distance of the center of the&  $0$  m \\ 
                     &   pinhole from the origin        &  \\ \hline
$h$             &   Planck's constant                   &   $6.62607004\times10^{-34}$  Js     \\    \hline
$\hbar$     &   Planck's constant/$2\pi$    &   $1.05457180\times10^{-34}$  Js \\ \hline
$m$ &         mass of electron                    &   $9.10938356*10^{-31}$     kg                    \\    \hline
$\alpha$ &          $\hbar/m$                     &   $0.00011576764$     Jsm$^{-1}$                    \\    \hline
$k_x$ &         Magnitude of $x$-wavenumber                     &   $1.295698717\times10^6$ m$^{-1}$                               \\    \hline
$k_y$ &        Magnitude of $y$-wavenumber                      &   $1.122938132\times10^{12}$   m$^{-1}$                  \\    \hline
$k_z$ &          Magnitude of $z$-wavenumber                       &   $0$                               \\    \hline
$v_x$ &          $x$-component of the velocity                   &   $\alpha k_x=150$ ms$^{-1}$                 \\      
            &          of the Gaussian wave-packet                     &                                                  \\    \hline
$v_y$ &          $y$-component of the velocity                   &   $\alpha k_y=1.3\times 10^8$   ms$^{-1}$                \\      
            &          of the Gaussian wave-packet                     &                                                  \\    \hline
$v_z$ &          $z$-component of the velocity                   &   $\alpha k_z=0$   ms$^{-1}$                \\      
            &          of the Gaussian wave-packet                      &                                                  \\    \hline
$\omega$ &   Angular frequency       $\omega$                     &   $\hbar(k_x^2+k_y^2)/2m$                         \\    \hline
$\chi$ &          phase shift of $\psi_2$                     &   $0$                         \\    \hline
$\Delta x_{n0}=\Delta z_{n0}$ &     Width of the $-x_0$  wave-packet      &   $\Delta x_{n0}=7\times 10^{-8}$  m    \\    \hline
$\Delta x_{p0}=\Delta z_{p0}$ &     Width of the $+x_0$ wave-packet      &   $\Delta x_{p0}=\Delta x_{n0}$      \\    \hline
$\Delta x_{n0}=\Delta z_{n0}$ &     Width of the $-x_0$  wave-packet     &    $\Delta x_{n0}=7\times 10^{-8}$   m      \\                                                                                     
                               &      for unequal pinhole widths  &                                                                                                               \\    \hline
$\Delta x_{p0}=\Delta z_{p0}$ &     Width of the $+x_0$ wave-packet    &         $\Delta x_{p0}=2\Delta x_{n0}$     \\                                                                                   
                              &        for unequal pinhole widths &                                                                                                                  \\    \hline
\end{tabular}
\end{center}
\newpage

The quantities used for the two-pinhole experiment plots are given in Table 3.1. Note that the slit width used by J\"{o}nsson is $2\times 10^{-7}$ m. The width of the Gaussian 
wavepackets $\psi_1$ and $\psi_2$ are defined at half their amplitude. We have chosen the widths of $\psi_1$ and $\psi_2$ to be $\Delta x_{n0}=\Delta z_{n0}=\Delta x_{p0}=\Delta z_{p0}=7\times 10^{-8}$ m so that the width of the base of $\psi_1$ and $\psi_2$ approximately corresponds to $2\times10^{-7}$ m. For unequal pinhole widths  $\Delta x_{p0}=\Delta z_{p0}=2\Delta x_{n0}$. We will consider three configurations: (1) Equal pinhole widths and equal amplitudes, (2) Equal pinhole widths and unequal amplitudes and (3) Unequal pinhole widths and equal amplitudes. 

In the two-dimensional case,  i.e., pinhole case,   the intensity $R^2$ and quantum potential $Q$ are functions of four variables $x,y,x$ and $t$. To produce plots we note that because the $y$-behaviour is represented by a plane-wave, $Q_y=0$, while $Q_x$ and $Q_z$ depend only on $x$, $z$ and $t$. Similarly, the intensity does not depend on $y$. This means that the values of $R^2$ and $Q$ in the  $xy$-plane at a given instant of time are the same from $y=-\infty$ to $y=+\infty$, as mentioned in \S \ref{MM}. At a later instant, the form of  $R^2$ and $Q$ change instantaneously from $y=-\infty$ to $y=+\infty$. This unphysical behaviour is due  to the use of the plane-wave idealisation to represent  the $y$-behaviour. A more realistic picture would be to also use a Gaussian  in the $y$-direction. However, as we shall see, the model produces a realistic picture of particle trajectories which depend on $x,y,z,t$.  Since the quantum potential and intensity change in time, the electron  `sees' evolving values of these quantities. All the plots below show what the electron `sees' at a particular instant of time $t$ and a particular position $x,y,z$.

To graphically represent $R^2$ and $Q$ we proceeded two ways. First, we produced animations of $R^2$ and $Q$.  We produced animations of six frames, so that the sequence of frames is short enough to be reproduced in this article. In any case,  our commuter did not have enough memory to produce animations of more than six frames. The animations are produced in the $xz$-plane and show the form of intensity and quantum potential that the electron `sees' at each instant of time as it moves along its trajectory. Second, we produced animations of density plots in the $xz$-plane. These results are presented  by  placing three two-dimensional  $xz$-slices (three frames of the animation) along the $t$-axis, i.e., we pick out three slices of a fully three dimensional density plot. 

\subsection{Computer plots for equal pinhole widths and equal amplitudes}

\begin{figure}[h]
\unitlength=1in 
\hspace*{1.1in}\includegraphics[width=4in,height=5.33in]  {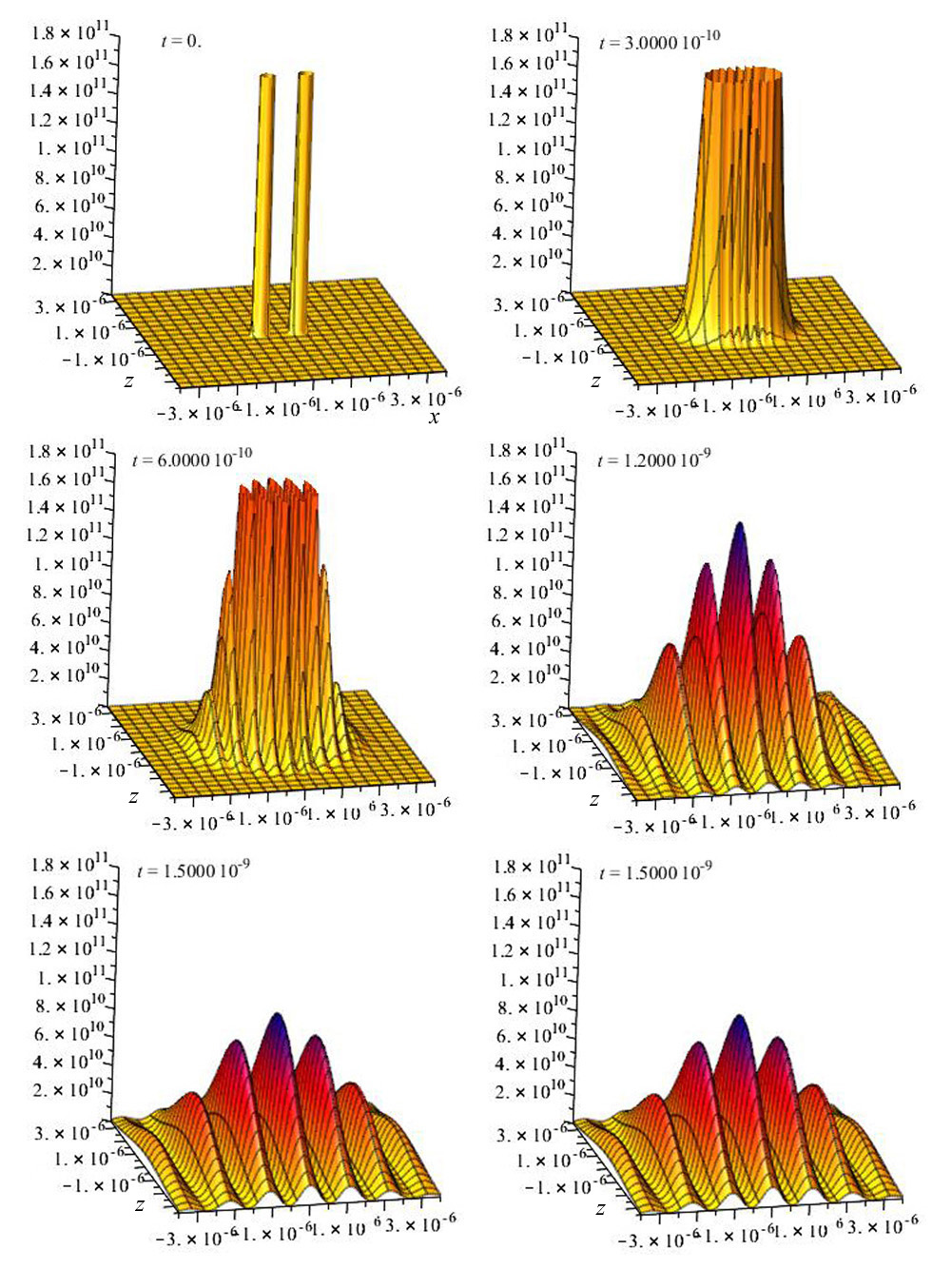} 
\caption{The intensity in a two-pinhole interference experiment with  equal widths and equal amplitudes modeled by two-dimensional Gaussian wave-packets with equal widths and equal amplitudes.\label{INT2DG}}
\end{figure}
The animation sequence for the intensity $R^2$ for equal widths and equal amplitudes (EQEA) is shown in Fig. \ref{INT2DG}. The animation ranges are $x=-3.5\times 10^{-6}$ to $3.5\times 10^{-6}$ m, $z=-3.5\times 10^{-6}$ to $3.5\times 10^{-6}$ m, $R^2 = 0$ to $1.8\times 10^{-11}$ Jm$^{-2}$s$^{-1}$ and $t=0$ to $1.5\times 10^{-9}$ s. The plots show the time evolution of the intensity in the $xz$-plane. The first frame shows the Gaussian peaks at the  two pinholes. Frames  two and three show the spreading Gaussian packets beginning to overlap and also show the beginning of the formation of interference fringes.   Frames four to six show the time evolution of distinct interference fringes. Frame six shows the intensity distribution at time $t=1.5\times 10^{-9}$ s which corresponds to a pinhole screen  to detecting screen separation of $y = 0.195$ m given that the  electron  velocity in the y-direction is $v_y = 1.3 \times 10^{-8}$ ms$^{-1}$. Our pinhole screen and detecting separation therefore differs from that in J\"{o}nssons experiment which was $0.35$ m corresponding a time evolution of $t=0$ to $2.6923\times 10^{-9}$ s. We chose this time in order to  show the beginnings of the overlap of the Gaussian wave packets. Using the J\"{o}nsson  time of $t=0$ to $2.6923\times 10^{-9}$ s resulted in a clear interference pattern in the second frame, missing out the early overlap.

We can make an approximate calculation of the visibility of the central fringe by taking readings from `face-on' plots, i.e., plots with the $xz$-plane  in the plane of the paper (not shown here). Readings can be taken from the plots shown by taking due consideration of the orientation, but even then, readings are less accurate than with face-on plots. Similarly, to calculate the visibility of the interference fringes for the case of unequal amplitudes and for the case of unequal widths, readings are taken from face-on plots not included here. The visibility of the central fringe for the EWEA case is:
\[
V_{EWEA}=\frac{I_{max}-I_{min}}{I_{max}+I_{min}}=\frac{10\times 10^{10}-0}{10\times 10^{10}+0}=1.
\]
\begin{figure}[h]
\unitlength=1in 
\hspace*{1.0in}\includegraphics[width=4in,height=3in]  {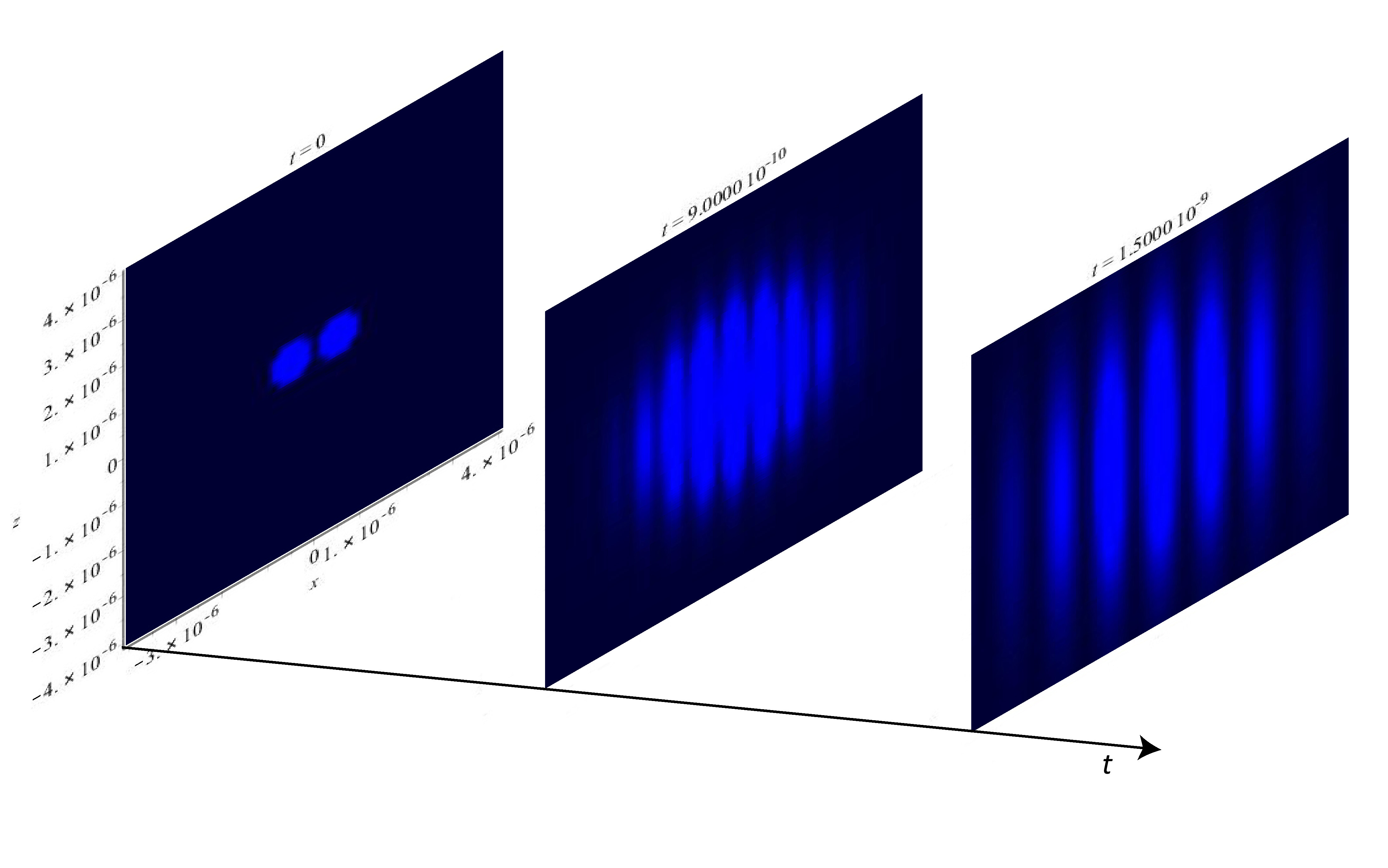} 
\caption{A sequence of density plots (3 slices of a 3D-plot) of the intensity $R^2$  in a two-pinhole interference experiment with  equal widths and equal amplitudes modeled by  two-dimensional Gaussian wave-packets with equal widths and equal amplitudes.\label{INT2DGDP}}
\end{figure}
A sequence of density plots (3 slices of a 3D-plot) of the intensity $R^2$ for equal widths and equal amplitudes is shown in Fig. \ref{INT2DGDP}. The plot  ranges are $x=-4\times 10^{-6}$ to $4\times 10^{-6}$ m, $z=-4\times 10^{-6}$ to $4\times 10^{-6}$ m  and $t=0$ to $1.5\times 10^{-9}$ s.  The first $xz$-slice shows the high intensity emerging from the two pinholes, the middle $xz$-slice shows the beginning of the formation of interference as the Gaussian packets begin to overlap, while the final $xz$-slice shows a fully formed interference pattern.
\begin{figure}[h]
\unitlength=1in
\hspace*{1.1in}\includegraphics[width=4in,height=5.33in]  {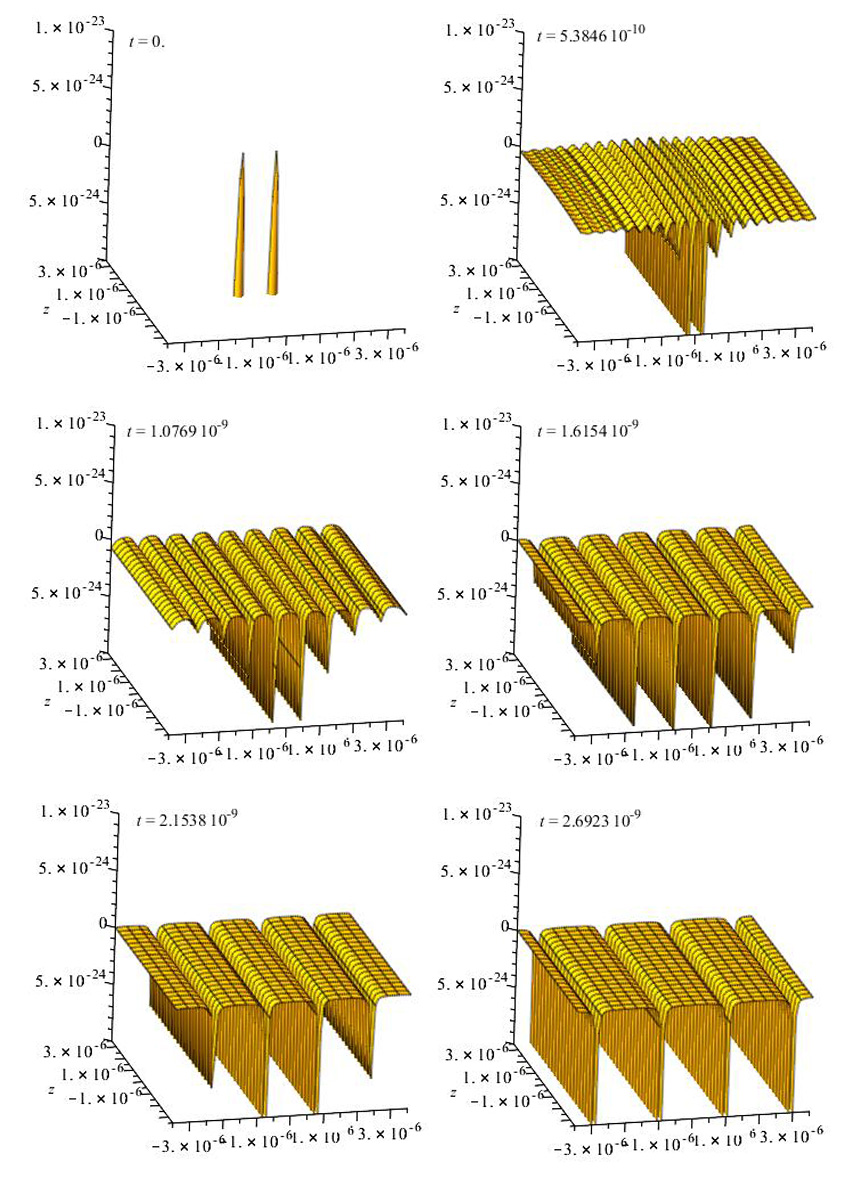}  
\caption{The quantum potential in a two-pinhole interference experiment with  equal widths and equal amplitudes modeled by two-dimensional Gaussian wave-packets with equal widths and equal amplitudes.\label{QPAnim}}
\end{figure}

The animation sequence for the quantum potential $Q$ for equal widths and equal amplitudes is shown in Fig. \ref{QPAnim}. The animation ranges are  $x=-3.5\times 10^{-6}$ to $3.5\times 10^{-6}$ m, $z=-3.5\times 10^{-6}$ to $3.5\times 10^{-6}$ m, $Q=-1\times 10^{-23}$ to $1\times 10^{-23}$ J and $t=0$ to $2.6923\times 10^{-9}$ s. This time we used the same pinhole to detecting screen  separation, 0.35 m, (corresponding to a time of flight of $t=2.6923\times10^{-9}$ s) as J\"{o}nsson, since this resulted in a clear plateau-valley formation in the final frame. The first frame shows that the quantum potential is restricted to the width of the two pinholes. The second frame shows the beginning of the formation of quantum potential plateaus and valleys  corresponding to the beginning of the overlap of the Gaussian wave packets. Subsequent frames show the continued widening of the plateaus and the deepening of the valleys. The final frame, as mentioned, shows clear plateau and valley formation. 
The gradient of the quantum potential gives rise to a quantum force. Where the gradient is zero, as on the flat plateaus, the quantum force is zero and electrons progress along their trajectory to a bright fringe on the detecting screen unhindered. At the edges of the plateaus the quantum potential slopes steeply down to the valleys. The steep gradient of these slopes gives rise to a large quantum force that pushes particles with trajectories along these slopes to adjacent plateaus, after which they proceed unhindered to a bright fringe on the detecting screen.  In this way, the quantum potential   guides the electrons to the bright fringes and prevents electrons reaching the dark fringes. Note though, as mentioned earlier, that since the $R$ and $S$-fields codetermine each other, the  $S$-field can also be said to guide the electrons to the bright fringes. 

\begin{figure}[h]
\unitlength=1in 
\hspace*{1.0in}\includegraphics[width=4in,height=3in]  {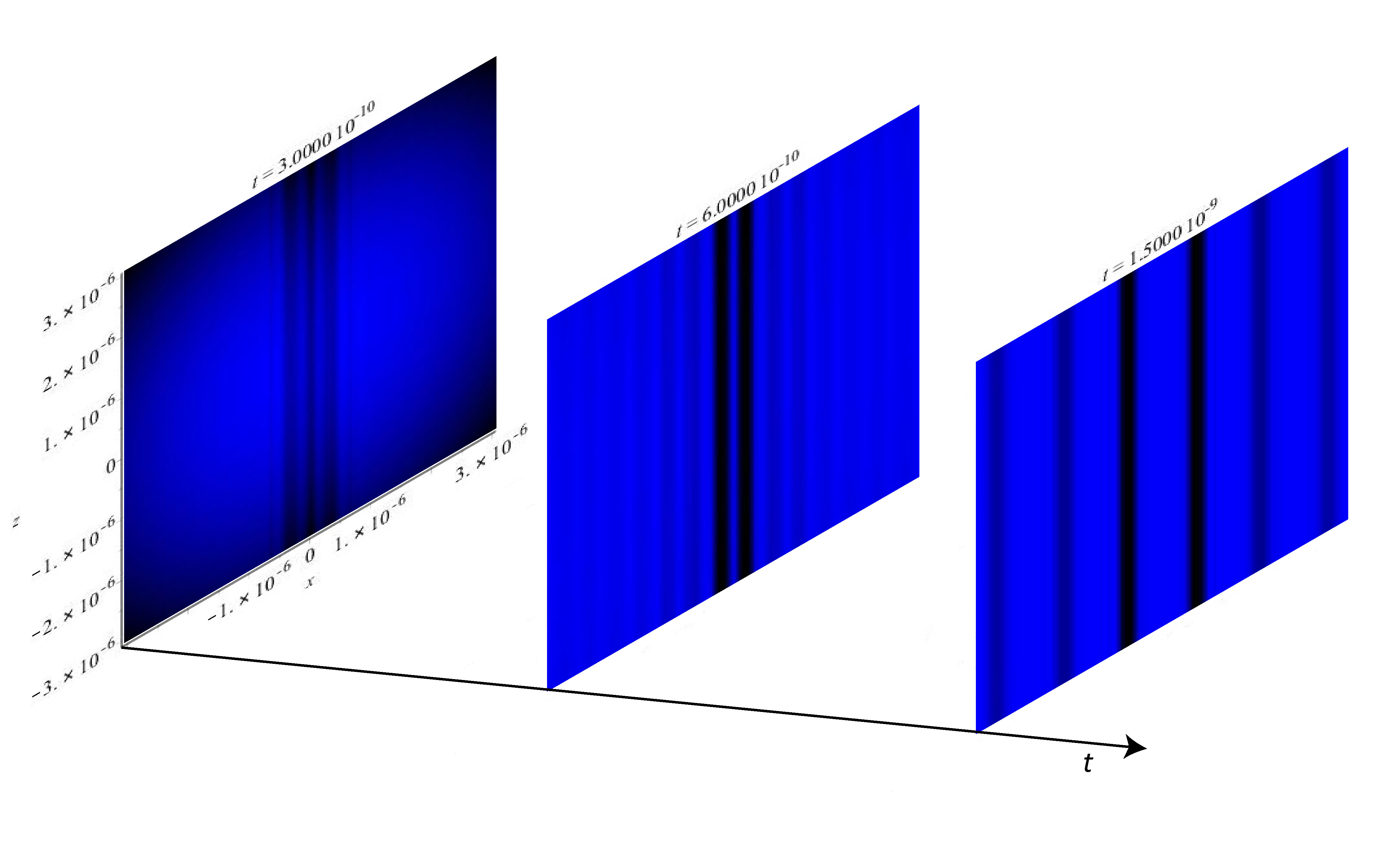} 
\caption{A sequence of density plots (3 slices of a 3D-plot) of the intensity $R^2$ in a two-pinhole interference experiment with  equal widths and equal amplitudes modeled by two-dimensional Gaussian wave-packets with equal widths and equal amplitudes.\label{T2DGQPDP}}
\end{figure}
A sequence of density plots (3 slices of a 3D-plot) of the quantum potential  for equal widths and equal amplitudes is shown in Fig. \ref{T2DGQPDP}. The plot  ranges are $x=-3\times 10^{-6}$ to $3\times 10^{-6}$ m, $z=-3\times 10^{-6}$ to $3\times 10^{-6}$ m  and $t=0$ to $1.5\times 10^{-9}$ s. When producing the density animation we found that part of the image in the $t=0$ s frame was missing, hence, we have left out this frame, beginning instead with the $t=3\times 10^{-10}$ s frame. The reason for the missing image  is not clear, but is most likely due to the density plotting algorithm not handling difficult numbers very well, unlike the 3-D plotting algorithm.  The first slice shows the beginning of the overlap of the Gaussian packets and the beginning of plateau and valley formation. The middle slice shows the more developed plateaus and valleys, while the final slice shows distinct plateaus and valleys. The wide bright blue bands indicate the quantum potential plateaus where the quantum force is zero. The narrower dark bands show the quantum potential sloping down to the valleys, slopes were electrons experience a strong quantum force. The darker the bands the steeper the quantum potential slopes.

\begin{figure}[h]
\unitlength=1in
\hspace*{0.8in}\includegraphics[width=4.5in,height=2.36in]  {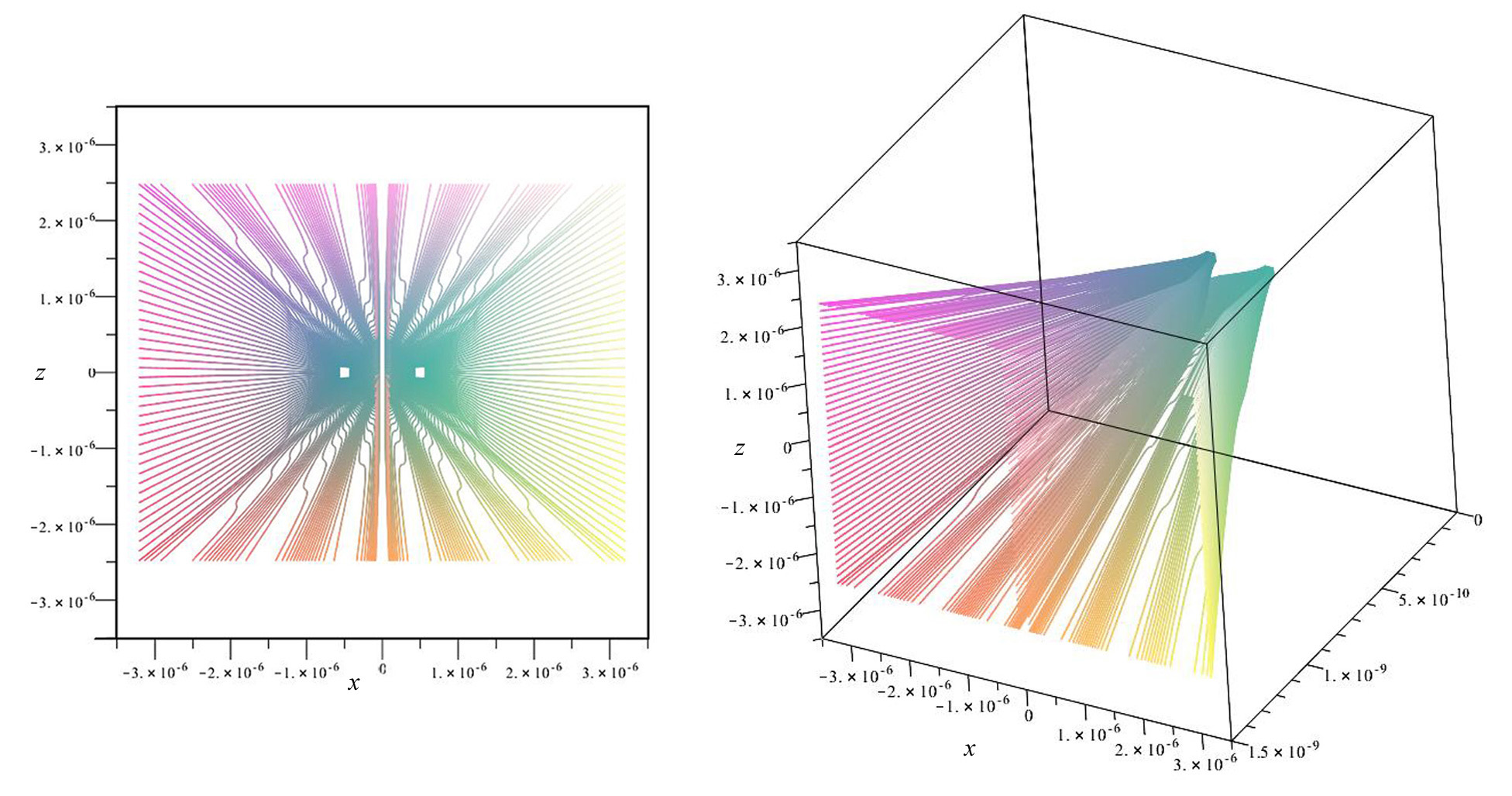}  
\caption{The trajectories in a two-pinhole interference experiment with  equal widths and equal amplitudes modeled by two-dimensional Gaussian wave-packets with equal widths and equal amplitudes.\label{TRAJEWEA}}
\end{figure}
The trajectories for equal widths and equal amplitudes is shown in Fig. \ref{TRAJEWEA}. The trajectory  ranges are $x=-3.5\times 10^{-6}$ to $3.5\times 10^{-6}$ m, $z=-3.5\times 10^{-6}$ to $3.5\times 10^{-6}$ m and $t=0$ to $1.5\times 10^{-9}$ s. Though the axes are labeled at the edges, the plots correspond to axes with their origin placed centrally between the pinholes. We have chosen to label the axis at the edges of the plot frame in order to show the trajectories clearly. In a real experiment, the initial position of the electrons can lie anywhere within the pinholes. But, to clearly show the behaviour of the trajectories we have chosen square initial positions within each pinhole. It is clear, that interference occurs only along the $x$-direction; there is no interference along the $z$-direction. We also see clearly how the quantum potential (and $S$-field)
guides the electron trajectories to the bright fringes. Electrons whose trajectories lie within the quantum potential plateaus, therefore experiencing no quantum force, move along straight trajectories to the bright fringes. Electrons whose trajectories lie along the quantum potential slopes are pushed by the quantum force to an adjacent plateau, thereafter proceeding along straight trajectories to the bright fringes.

\subsection{Computer plots for equal pinhole widths and unequal amplitudes}

\begin{figure}[h]
\unitlength=1in 
\hspace*{1.1in}\includegraphics[width=4in,height=5.33in]  {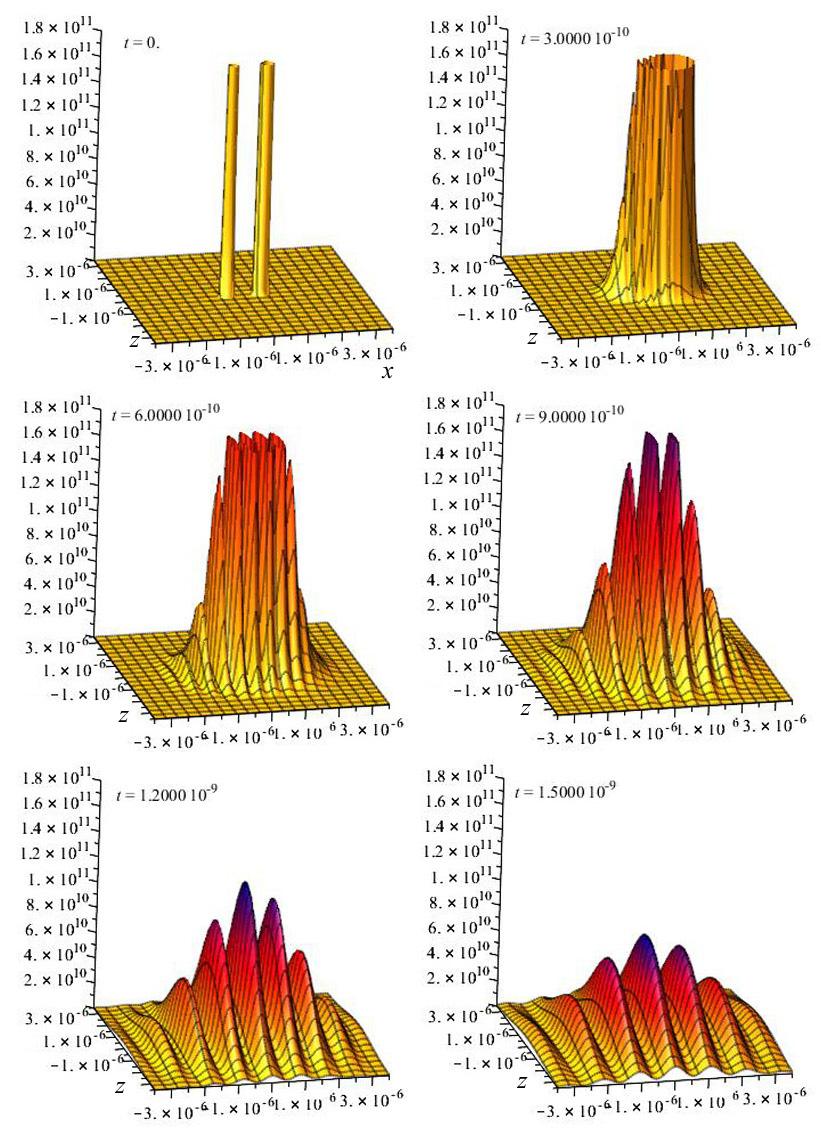} 
\caption{The intensity in a two-pinhole interference experiment with  equal widths but unequal amplitudes modeled by  two-dimensional Gaussian wave-packets with  equal widths but unequal amplitudes.\label{INT2DGUA}}
\end{figure}
The animation sequence for the intensity $R^2$ for equal widths and unequal amplitudes (EWUA) is shown in Fig. \ref{INT2DGUA}. The animation ranges are $x=-3.5\times 10^{-6}$ to $3.5\times 10^{-6}$ m, $z=-3.5\times 10^{-6}$ to $3.5\times 10^{-6}$ m, $R^2 = 0$ to $1.8\times 10^{-11}$ Jm$^{-2}$s$^{-1}$ and $t=0$ to $1.5\times 10^{-9}$ s. As indicated in Table 3.1, for the case  EWUA the angle  $b=\frac{\pi}{3}$. This results in an increase in the intensity through  the $+x_0$-pinhole from $\frac{1}{2}$ to $\frac{3}{4}$ and a reduction in the intensity at the $-x_0$-pinhole from $\frac{1}{2}$ to $\frac{1}{4}$. From Fig. \ref{INT2DGUA}, we see that as the Gaussian wave-packets begin to combine to form a single peak envelope with interference fringes beginning to form, the intensity peak is shifted toward the larger intensity $+x_0$-pinhole. This shift becomes less pronounced, almost disappearing,  as the  interference fringes become more distinct as in the last  $t=0$ to $1.5\times 10^{-9}$ s frame. Comparing the $t=1.5\times10^{-9}$ frame for EWUA with the corresponding intensity frame for EWEA we can see visually that fringe visibility is reduced. We can confirm this visual observation by calculating the visibility of the central fringe:
\[
V_{EWUA}=\frac{I_{max}-I_{min}}{I_{max}+I_{min}}=\frac{8\times 10^{10}-2\times 1^{10}}{8\times 10^{10}+2\times 1^{10}}=0.6
\]
Clearly, the visibility is lower for the EWUA case.


\begin{figure}[h]
\unitlength=1in 
\hspace*{1.0in}\includegraphics[width=4in,height=3in]  {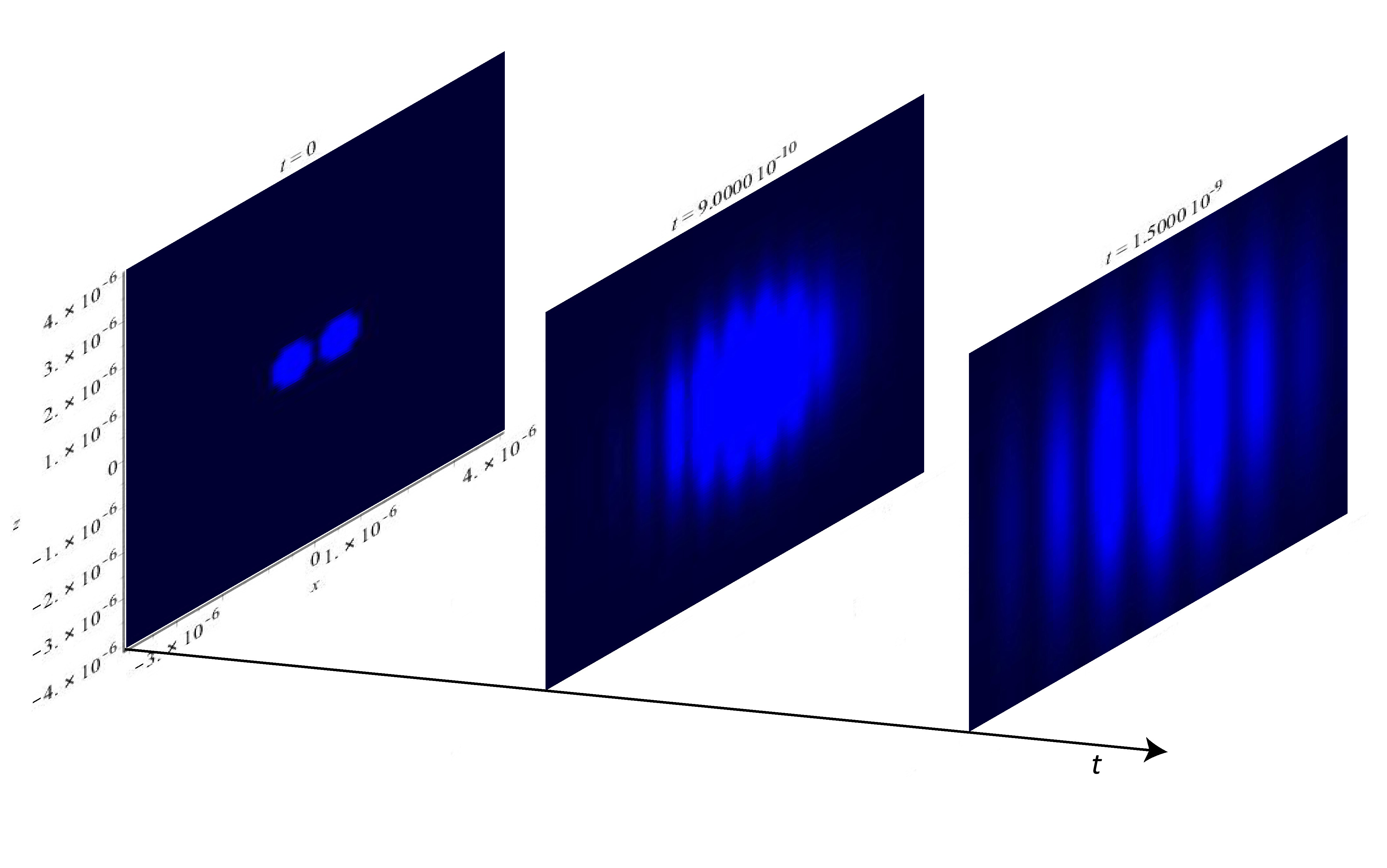} 
\caption{A sequence of density plots (3 slices of a 3D-plot) of the intensity $R^2$  in a two-pinhole interference experiment with  equal widths but unequal amplitudes modeled by  two-dimensional Gaussian wave-packets with equal widths but unequal amplitudes.\label{INT2DGDPUA}}
\end{figure}
A sequence of density plots (3 slices of a 3D-plot) of the intensity $R^2$ for equal widths and unequal amplitudes is shown in  Fig. \ref{INT2DGDPUA}. The plot  ranges are $x=-4\times 10^{-6}$ to $4\times 10^{-6}$ m, $z=-4\times 10^{-6}$ to $4\times 10^{-6}$ m  and $t=0$ to $1.5\times 10^{-9}$ s. Again, comparing with EWUA case, we see that the intensity is reduced by noticing that the dark bands are not as distinct as for the case EWEA.

\begin{figure}[h]
\unitlength=1in
\hspace*{1.1in}\includegraphics[width=4in,height=5.33in]  {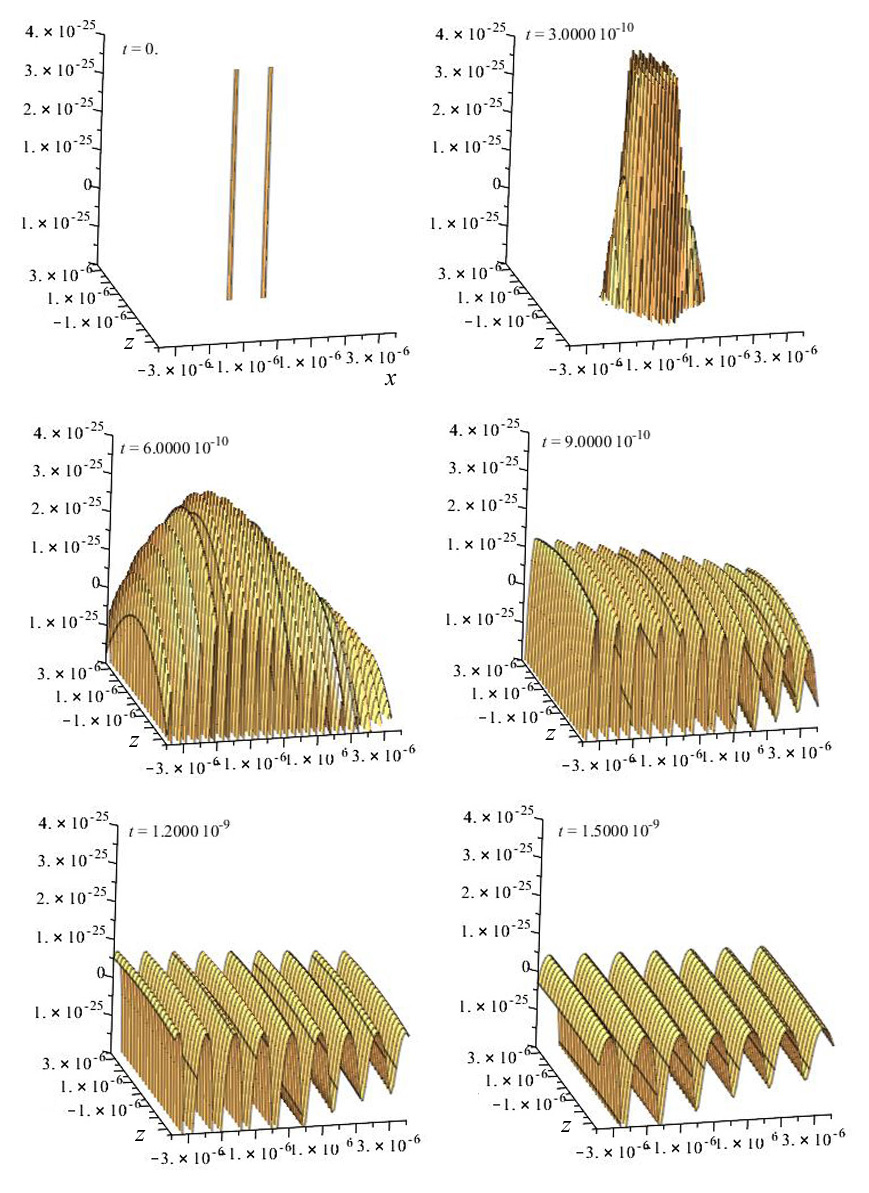}  
\caption{The quantum potential in a two-pinhole interference experiment with  equal widths but unequal amplitudes modeled by  two-dimensional Gaussian wave-packets with equal widths but unequal amplitudes.\label{QPAnimUA}}
\end{figure}
The animation sequence for the quantum potential $Q$ for equal widths and unequal amplitudes is shown in Fig.  \ref{QPAnimUA}. The animation ranges are  $x=-3.5\times 10^{-6}$ to $3.5\times 10^{-6}$ m, $z=-3.5\times 10^{-6}$ to $3.5\times 10^{-6}$ m, $Q=-2\times 10^{-25}$ to $4\times 10^{-25}$ J and $t=0$ to $1.5\times 10^{-9}$ s. The quantum potential in the early frames, perhaps unexpectedly, peaks on the side of the  lower intensity  $-x_0$-pinhole. This behaviour is most pronounced in frames two and three.  As the peaks and valleys become more pronounced, the envelope  peak spreads and flattens as shown in frames 5 and 6. However, the valleys on the side of the lower intensity $-x_0$-pinhole are deeper. Correspondingly, the gradient of quantum potential sloping down to the deeper valleys is greater giving rise to a stronger quantum force. This results in the formation of more distinct fringes  on the side of the lower intensity pinhole.This feature is hardly visible in either the intensity animation frames, Fig.  \ref{INT2DGUA}, or in the intensity density plots, Fig. \ref{INT2DGDPUA}. However, as we shall see below, the trajectory plots, Fig. \ref{TRAJEWUA}, shows this feature more clearly.


\begin{figure}[h]
\unitlength=1in 
\hspace*{1.0in}\includegraphics[width=4in,height=3in]  {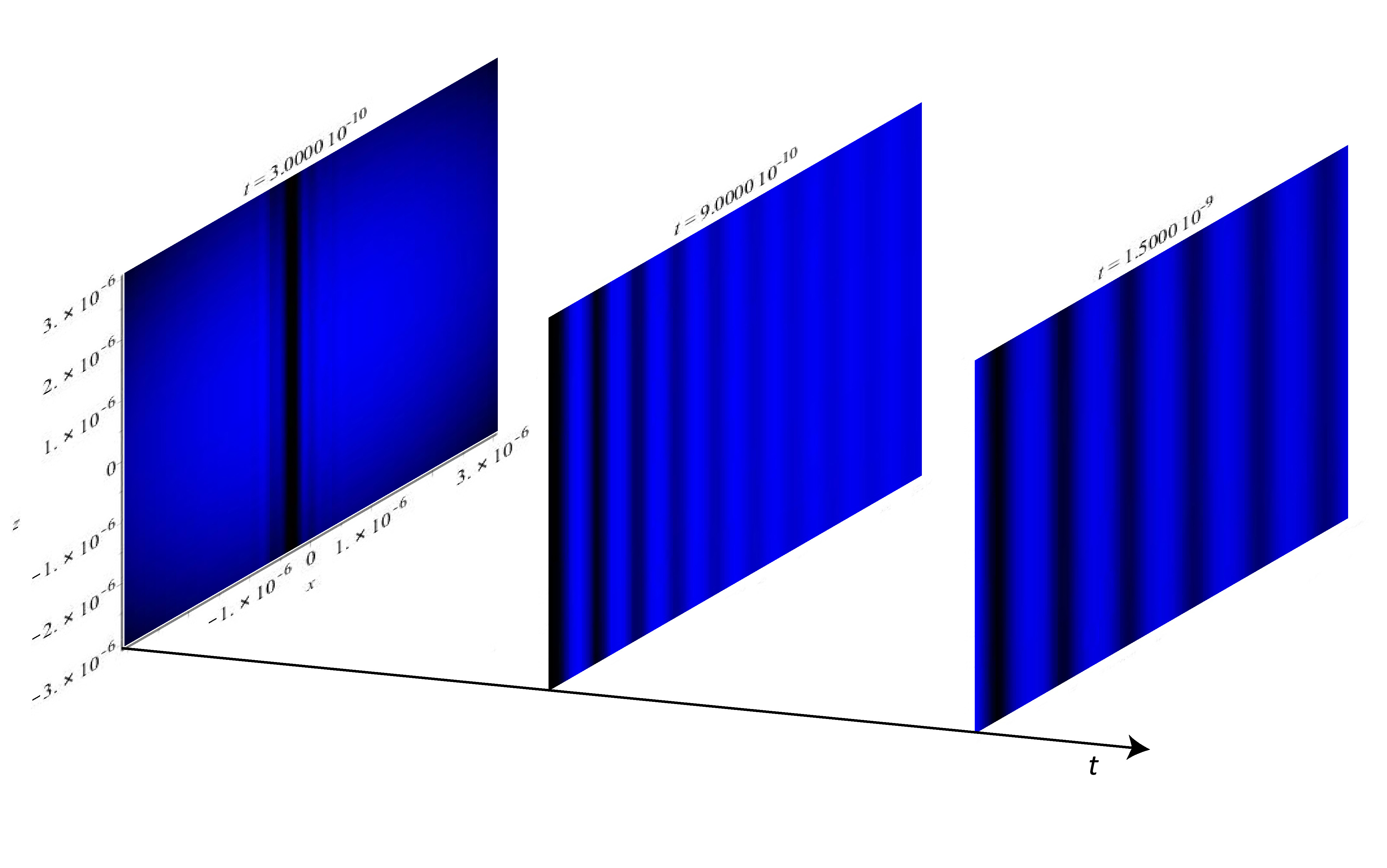} 
\caption{A sequence of density plots (3 slices of a 3D-plot) of the intensity $R^2$ in a two-pinhole interference experiment with  equal widths but unequal amplitudes modeled by two-dimensional Gaussian wave-packets with equal widths but unequal amplitudes.\label{T2DGQPDPUA}}
\end{figure}
A sequence of density plots (3 slices of a 3D-plot) of the quantum potential  for equal widths and unequal amplitudes is shown in Fig.   \ref{T2DGQPDPUA}. The plot  ranges are $x=-3\times 10^{-6}$ to $3\times 10^{-6}$ m, $z=-3\times 10^{-6}$ to $3\times 10^{-6}$ m  and $t=0$ to $1.5\times 10^{-9}$ s. As for the EWEA case, the first $t=0$ s slice is not shown, this time, because the image it is badly formed. Instead, we begin with the  $t=3\times 10^{-9}$ s slice.  This slice clearly shows that the deeper valleys, indicated by blacker bands, are on the side of lower intensity $-x_0$-pinhole. As above, the bright blue bands represent regions where the quantum potential gradient is either zero or very small,  giving rise  to either a zero or small quantum force. In the final $t=1.5\times 10^{-9}$ s slice, the peaks and  valleys even out though a slight bias to deeper valleys on the $-x_0$ side is still discernible.

\begin{figure}[h]
\unitlength=1in
\hspace*{0.8in}\includegraphics[width=4.5in,height=2.36in]  {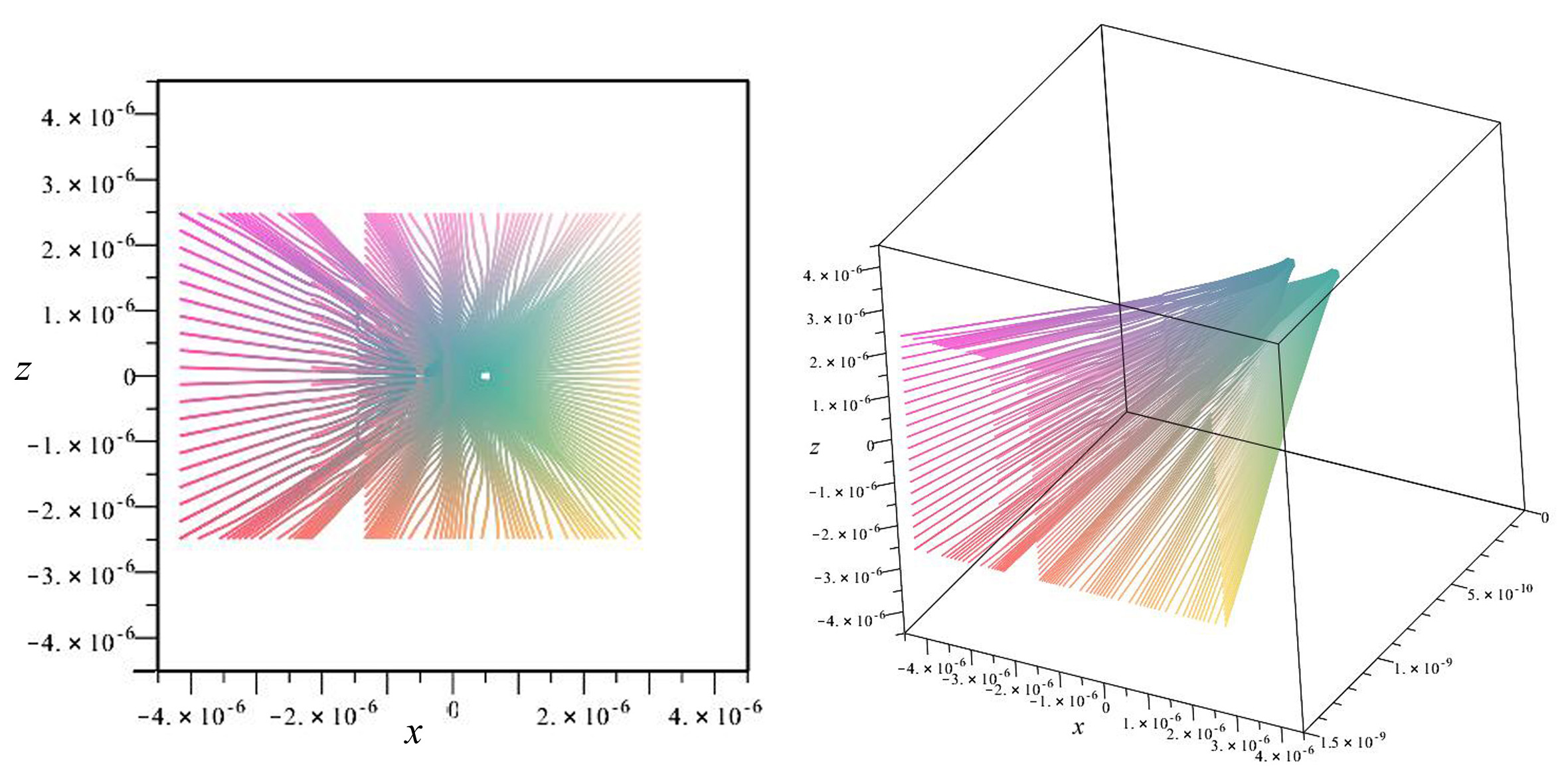}  
\caption{The trajectories in a two-pinhole interference experiment with  equal widths but unequal amplitudes modeled by two-dimensional Gaussian wave-packets with equal widths but unequal amplitudes.\label{TRAJEWUA}}
\end{figure}
The trajectories for equal widths and unequal amplitudes are shown in Fig.  \ref{TRAJEWUA}. The trajectory  ranges are $x=-4.5\times 10^{-6}$ to $4.5\times 10^{-6}$ m, $z=-4.5\times 10^{-6}$ to $4.5\times 10^{-6}$ m and $t=0$ to $1.5\times 10^{-9}$ s. We notice that some electron trajectories reach what were the dark regions for the case of EWEA. This indicates the reduction of fringe visibility that we saw above in the intensity plots for this case. We also notice that this reduced intensity  is less pronounced on the side of the  lower intensity $-x_0$-pinhole, so that interference fringes on this side are more distinct, a feature we noted above for the quantum potential for this case. The overall reduction in visibility is clear to see.

\subsection{Computer plots for unequal pinhole widths and equal amplitudes}
\begin{figure}[h]
\unitlength=1in 
\hspace*{1.1in}\includegraphics[width=4in,height=5.33in]  {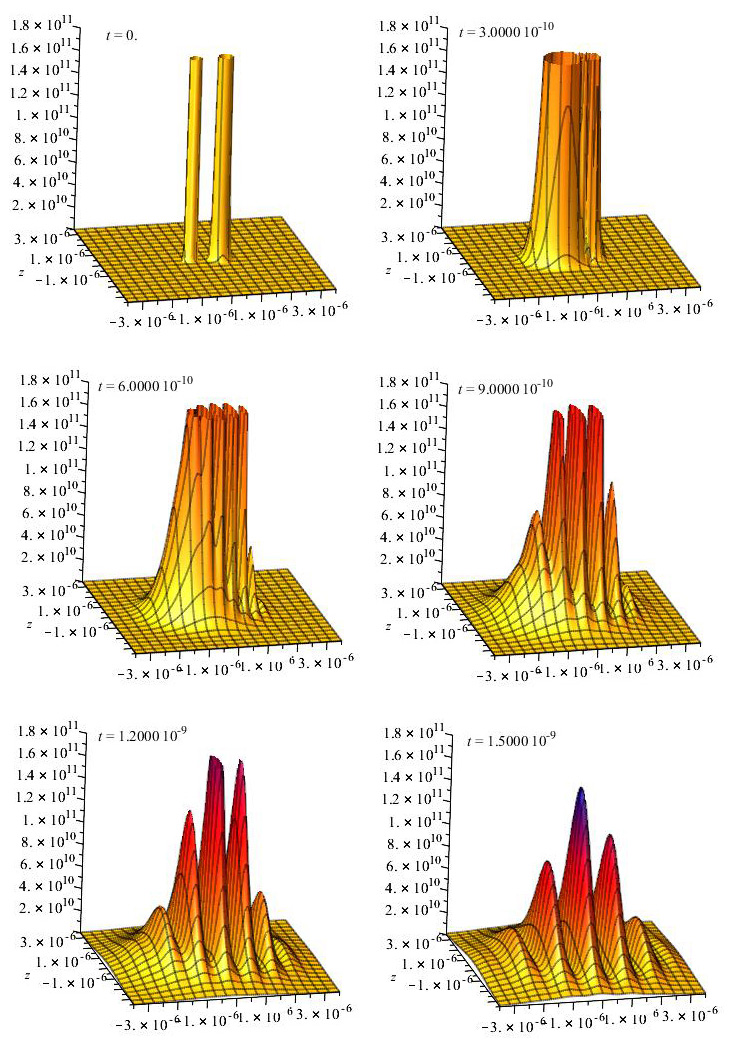} 
\caption{The intensity in a two-pinhole interference experiment with unequal widths but equal amplitudes modeled by two-dimensional Gaussian wave-packets with unequal widths but equal amplitudes.\label{INT2DGUW}}
\end{figure}
The animation sequence for the intensity $R^2$ for unequal widths and equal amplitudes is shown in Fig.  \ref{INT2DGUW}. The animation ranges are $x=-3.5\times 10^{-6}$ to $3.5\times 10^{-6}$ m, $z=-3.5\times 10^{-6}$ to $3.5\times 10^{-6}$ m, $R^2 = 0$ to $1.8\times 10^{11}$ Jm$^{-2}$s$^{-1}$ and $t=0$ to $1.5\times 10^{-9}$ s. From Table 3.1, we note that the width of the $+x_0$ Gaussian wave-packet is twice that of the  $-x_0$ Gaussian wave-packet. The first frame clearly shows the narrower  $-x_0$ wave-packet. A narrower wave-packet spreads more rapidly than a wider packet, as seen in the second frame. The more rapid spread of the narrower wave-packet results in the   wave-packets beginning  to overlap  on $+x$-side, as is shown in the second frame. As the wave-packets spread, the interference pattern becomes ever more distinct. Though becoming  a little more symmetrical about $x=0$, the fringe pattern is shifted  toward the $+x$-side, with the  fringes on the $+x$-side being slightly more pronounced.  This feature  is seen more clearly in the intensity density plots, which we will describe next. The visibility of the central fringe for this case is:
\[
V_{UWEA}=\frac{I_{max}-I_{min}}{I_{max}+I_{min}}=\frac{1.6\times 10^{11}-0.1\times 1^{11}}{1.6\times 10^{11}+0.1\times 10^{11}}=0.88.
\]
We see that the visibility is less than in the EWEA case, but greater than in the EWUA case.

\begin{figure}[h]
\unitlength=1in 
\hspace*{1.0in}\includegraphics[width=4in,height=3in]  {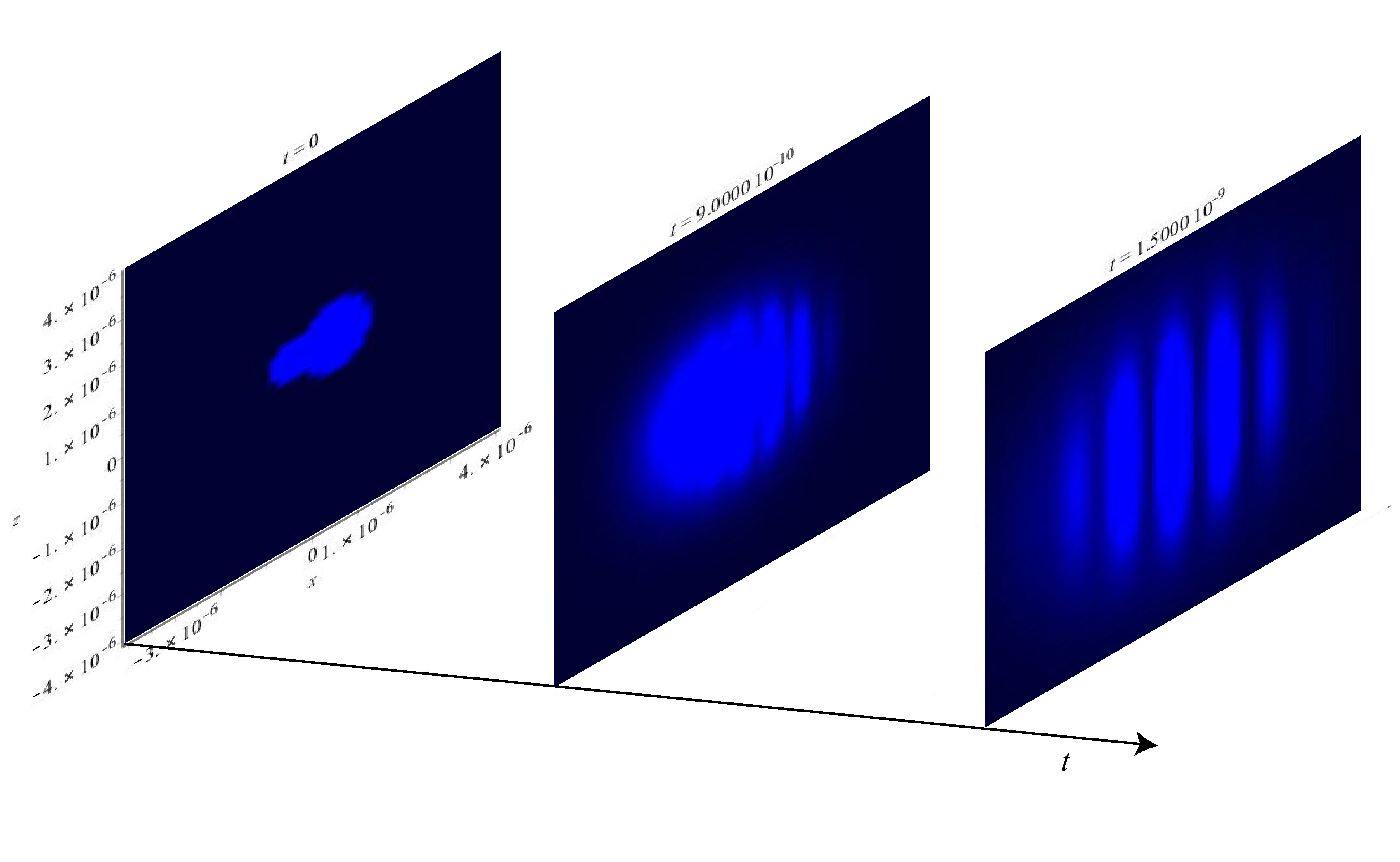} 
\caption{A sequence of density plots (3 slices of a 3D-plot) of the intensity $R^2$ in a two-pinhole interference experiment  with unequal widths and equal amplitudes  modeled by two-dimensional Gaussian wave-packets with unequal widths but equal amplitudes.\label{INT2DGDPUW}}
\end{figure}
A sequence of density plots (3 slices of a 3D-plot) of the intensity $R^2$ for unequal widths and equal amplitudes is shown in Fig. \ref{INT2DGDPUW}. The plot  ranges are $x=-4\times 10^{-6}$ to $4\times 10^{-6}$ m, $z=-4\times 10^{-6}$ to $4\times 10^{-6}$ m  and $t=0$ to $1.5\times 10^{-9}$ s. Again, the first slice clearly shows the difference in the  size of the pinholes, while the second slice  shows interference fringes beginning to  form on the $+x$-side. The final slice shows that the interference pattern develops into a more symmetric form, though still shifted more to the $+x$-side  and slightly more  pronounced on this side. The dark bands are less distinct than in the EWEA case, reflecting the reduction in fringe visibility for this case. 

\begin{figure}[h]
\unitlength=1in
\hspace*{1.1in}\includegraphics[width=4in,height=5.33in]  {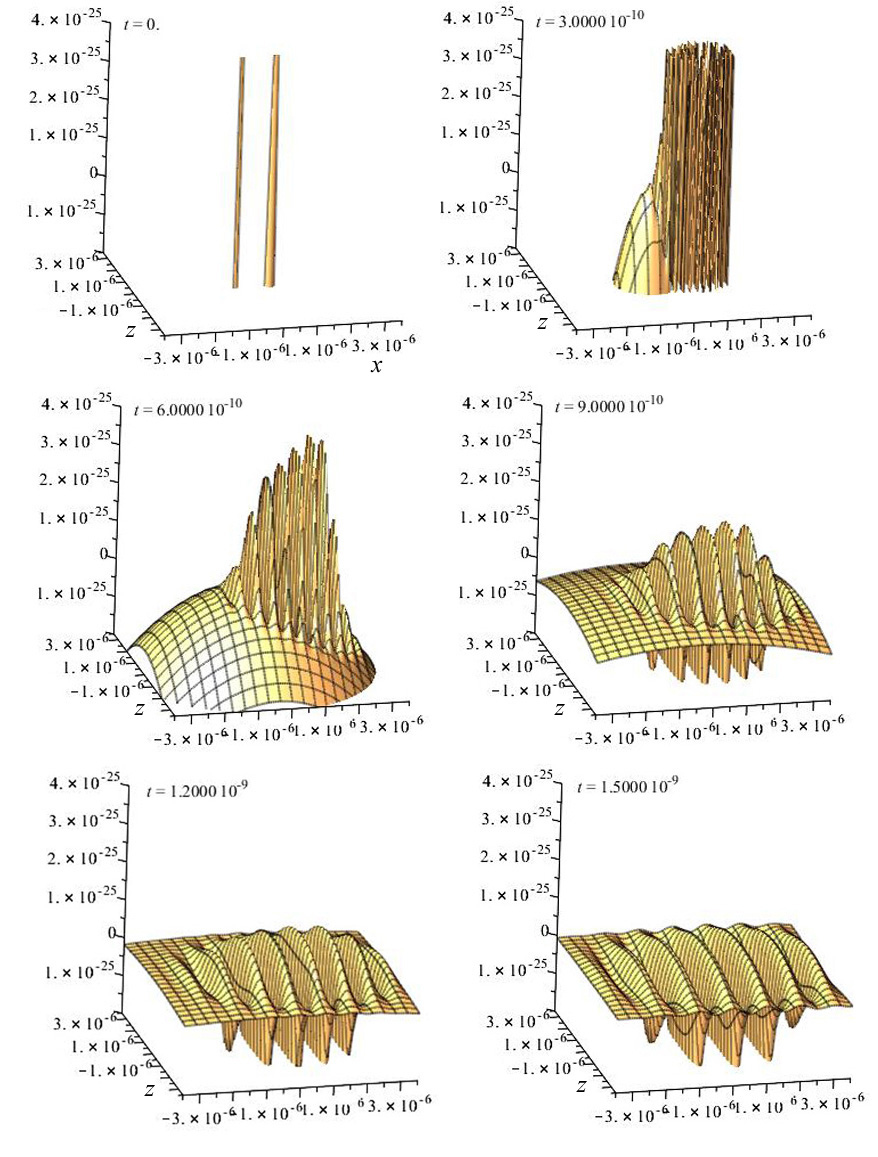}  
\caption{The quantum potential in a two-pinhole interference experiment with unequal widths but equal amplitudes modeled by two-dimensional Gaussian wave-packets with unequal widths but equal amplitudes.\label{QPAnimUW}}
\end{figure}
The animation sequence for the quantum potential $Q$ for unequal widths and equal amplitudes is shown in Fig.  \ref{QPAnimUW}. The animation ranges are  $x=-3.5\times 10^{-6}$ to $3.5\times 10^{-6}$ m, $z=-3.5\times 10^{-6}$ to $3.5\times 10^{-6}$ m, $Q=-2\times 10^{-25}$ to $4\times 10^{-25}$ J and $t=0$ to $1.5\times 10^{-9}$ s. As for the  interference animation, the first frame shows the difference in size of the pinholes, while the second frame shows the more  rapid spread of the narrower wave-packet. The overlap of the wave-packets, as for the intensity, begins on the $+x$-side, as does the formation of plateaus and valleys. Subsequent frames show the skewed formation to the $+x$-side of plateaus and valleys. The shift to the $+x$-side is maintained even in the last frame, even though the pattern looks more symmetrical. This can be seen by noticing that in the last frame there are three quantum potential peaks on the  $+x$-side, compared to two peaks on the $-x$-side.

\begin{figure}[h]
\unitlength=1in 
\hspace*{1.0in}\includegraphics[width=4in,height=3in]  {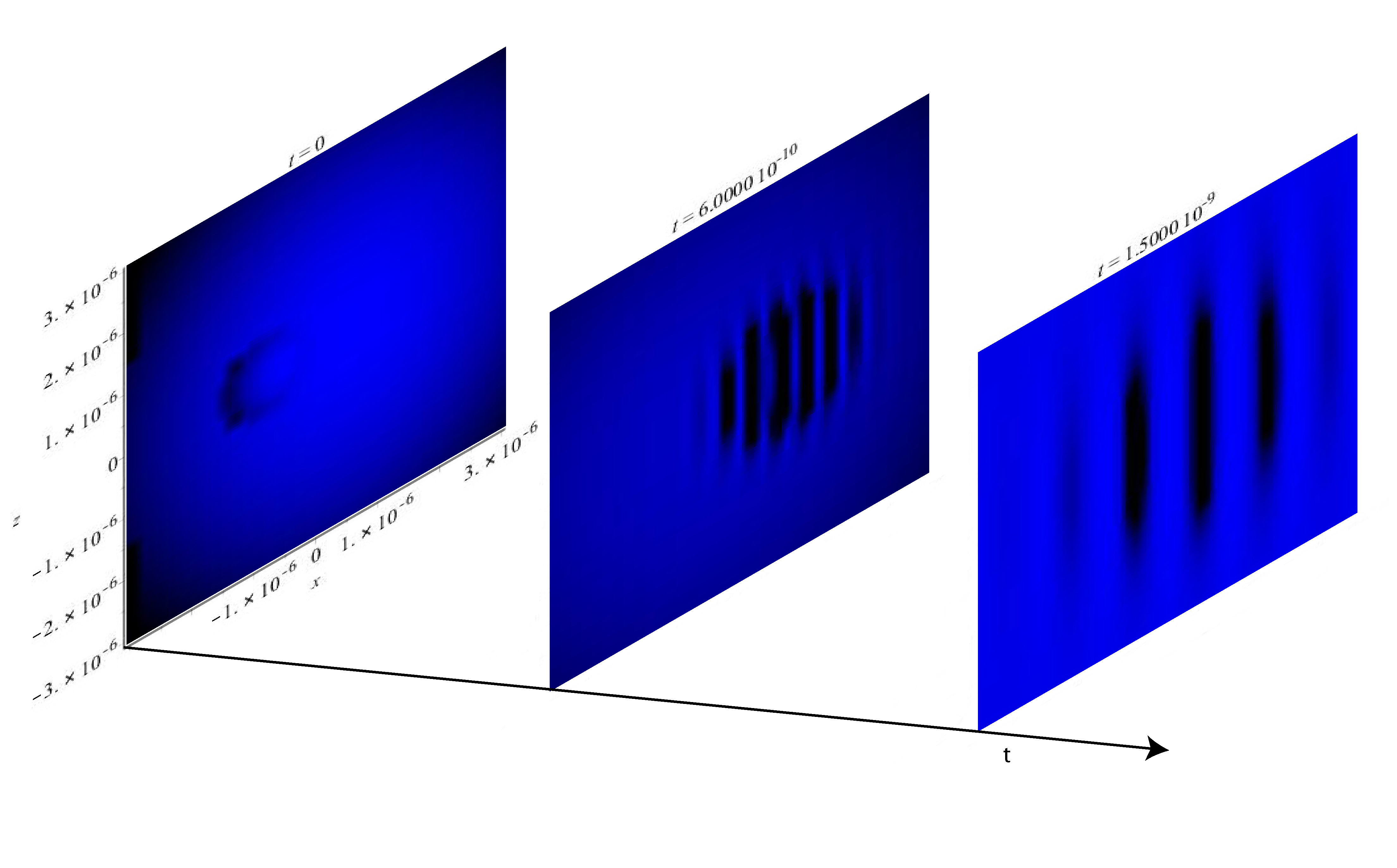} 
\caption{A sequence of density plots (3 slices of a 3D-plot) of the quantum potential  in a two-pinhole interference experiment with unequal widths but equal amplitudes modeled by two-dimensional Gaussian wave-packets with unequal widths but equal amplitudes.\label{T2DGQPDPUW}}
\end{figure}
A sequence of density plots (3 slices of a 3D-plot) of the quantum potential  for unequal widths and equal amplitudes is shown in Fig.  \ref{T2DGQPDPUW}. The plot  ranges are $x=-3\times 10^{-6}$ to $3\times 10^{-6}$ m, $z=-3\times 10^{-6}$ to $3\times 10^{-6}$ m  and $t=0$ to $1.5\times 10^{-9}$ s. As with the quantum potential density plots for the EWUA case,  the image in the first slice is problematic. This time, the image is complete but it does not reflect the two-pinhole structure.  As before, this is probably because the density plotting algorithm does not handle difficult numbers very well. The middle slice shows the early formation of plateaus and valleys skewed to  the $+x$-side,  with the plateau regions narrower than the valley regions. In the final slice, the plateau regions become wider than the valley regions, but are less distinct as compared to  the EWEA case, or even as compared to the EWUA case. Despite this, the fringe visibility is greater than for the EWUA case, though of course, less than for EWEA case, as we saw above. Again, the shift of quantum potential peaks to the $+x$-side can be seen.

\begin{figure}[h]
\unitlength=1in
\hspace*{0.8in}\includegraphics[width=4.5in,height=2.36in]  {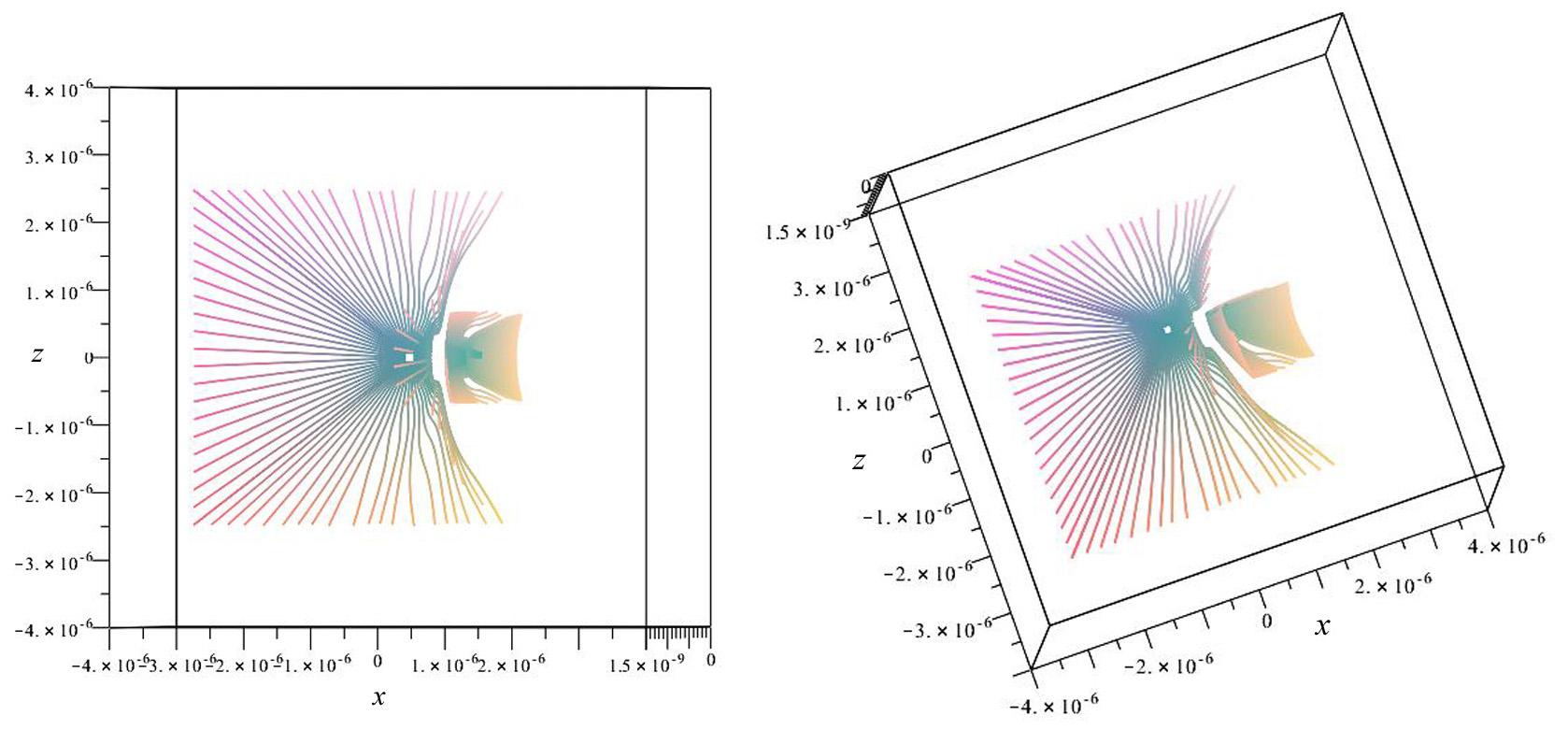}  
\caption{The trajectories in a two-pinhole interference experiment with unequal widths but equal amplitudes modeled by two-dimensional Gaussian wave-packets with  unequal widths but equal amplitudes.\label{TRAJEWEAUW}}
\end{figure}
The trajectories for unequal widths and equal amplitudes is shown in Fig.  \ref{TRAJEWEAUW}. The trajectory  ranges are $x=-4\times 10^{-6}$ to $4\times 10^{-6}$ m, $z=-4\times 10^{-6}$ to $4\times 10^{-6}$ m and $t=0$ to $1.5\times 10^{-9}$ s. The electron trajectories clearly show the rapid spread of the narrower $-x_0$-Gaussian wave-packet. It can also be seen that electron trajectories from the $-x_0$-pinhole are more evenly spread on the detecting screen parallel to the $x$-axis than for the EWEA case, indicating a reduced interference pattern. The electron trajectories from the  $+x_0$-pinhole, spread much less. Though only one bright and one dark fringe is shown on the $+x$-side, they appear more distinct than on the $-x$-side, reflecting the shift of the interference fringes to the $+x$-side
\section{Conclusion}
We have seen that the behaviour of the intensity, quantum potential and electron trajectories is similar to that for the  two-slit experiment modeled by  one-dimensional Gaussian wave-packets. In particular, the distinctive kinked behaviour of the electron trajectories is seen in directions parallel to the $x$-axis.  We saw in addition, as could possibly be guessed at the outset, that there is no interference in the vertical $z$-direction. We also saw the expected reduction in interference for the cases of unequal amplitudes and unequal widths.The reduction in interference  is interpreted, as in the classical case, in terms of  wave profiles with reduced coherence. This is a far more intuitive explanation for the reduction in fringe visibility (and in my view more appealing) than the common interpretation based on the Wootters-Zureck version of complementarity \cite{WZ79}, where the reduction  of interference for the cases of unequal amplitudes  and unequal widths   is attributed to partial particle behaviour and partial wave-behaviour. The partial particle behaviour is attributed to the increase in knowledge of the electrons path in the sense that an electron it is more likely to pass through the larger pinhole or the pinhole with the larger intensity. In references \cite{K92} and \cite{K2016} we argued that the Wootters-Zureck version of complementarity  as commonly interpreted actually contradicts Bohr's principle of complementary. In reference \cite{K2016} we also indicated that by   reference to two future, mutually exclusive experimental  arrangements, an interpretation of the Wootters-Zureck version of complementarity consistent with Bohr's principle of complementarity can be achieved.

Using the weak measurement protocol introduced by Aharanov, Albert and Vaidman (see reference \cite{K2017} for a brief overview and further references), Kocsis et al reproduced experimentally Bohm's trajectories in a two-slit interference experiment \cite{KOCSIS2011}. We might guess that it would not be difficult to modify the experiment slightly to reproduce the electron trajectories calculated here for the case of unequal widths and the case of unequal amplitudes.

\end{document}